\documentclass[fleqn,10pt,table,xcdraw]{wlscirep}
\usepackage[utf8]{inputenc}
\usepackage[T1]{fontenc}

\usepackage{float} % to force the placement of figures at specific locations (https://tex.stackexchange.com/questions/8625/force-figure-placement-in-text)
\makeatletter
\@fleqnfalse
\@mathmargin\@centering
\makeatother

\usepackage{fancyhdr,graphicx,amsmath,amssymb, amsfonts,mathtools}
\usepackage{euscript} %euler script
\usepackage{booktabs}
\usepackage{makecell}
\usepackage{amsmath}
\usepackage{amssymb}
\usepackage[T2A,T1]{fontenc}
\DeclareSymbolFont{cyrillic}{T2A}{cmr}{m}{n}%{it}
\SetSymbolFont{cyrillic}{bold}{T2A}{cmr}{bx}{n}%{it}
\DeclareMathSymbol{\mBe}{\mathord}{cyrillic}{193}%Б б 225
\DeclareMathSymbol{\mDe}{\mathord}{cyrillic}{196}%Д д 228
\DeclareMathSymbol{\mZhe}{\mathord}{cyrillic}{198}%Ж ж 230
\DeclareMathSymbol{\mI}{\mathord}{cyrillic}{200}%И и
\DeclareMathSymbol{\mEl}{\mathord}{cyrillic}{203}%Л л 235
\DeclareMathSymbol{\mTse}{\mathord}{cyrillic}{214}%Ц ц 246 or 84 116 ?
\DeclareMathSymbol{\mChe}{\mathord}{cyrillic}{215}%Ч ч 247
\DeclareMathSymbol{\mSha}{\mathord}{cyrillic}{216}%Ш ш 248
\DeclareMathSymbol{\mE}{\mathord}{cyrillic}{221}%Э э 253
\DeclareMathSymbol{\mYu}{\mathord}{cyrillic}{222}%Ю ю 254
\DeclareMathSymbol{\mYa}{\mathord}{cyrillic}{223}%Я я 255
\usepackage{academicons}
\usepackage{xcolor}
\usepackage{hyperref}
\definecolor{orcidlogocol}{HTML}{A6CE39}
\usepackage[
	linesnumbered,
	boxed,
	ruled,
	vlined
]{algorithm2e}
\hypersetup{
    colorlinks=true,
    linkcolor=blue,
    filecolor=magenta,      
    urlcolor=cyan,
}
\usepackage{graphicx}
\graphicspath{ {./img/} }

\SetKwComment{Comment}{/* }{ */}
\title{EquiCity Game: A mathematical serious game for participatory design of spatial configurations}
\newcommand{\orcid}[1]{\href{https://orcid.org/#1}{\textcolor[HTML]{A6CE39}{\aiOrcid}}}

\author[1,*]{Pirouz Nourian}% \orcid{0000-0002-3817-7931}}
\author[2]{Shervin Azadi}
\author[3]{Nan Bai}
\author[4]{Bruno de Andrade}
\author[5]{Nour Abu Zaid}
\author[6]{Samaneh Rezvani}
\author[3]{Ana Pereira Roders}
\affil[1]{University of Twente, Department of Planning and Geoinformation Management, Enschede, 7522 NH, Netherlands}
\affil[2]{Eindhoven University of Technology, Department of Built Environment, Eindhoven, 5612 AZ, Netherlands}
\affil[3]{Delft University of Technology, Department of Engineering and Technology, Delft, 2628 BL, Netherlands}

\affil[4]{University Portucalense, Department of Architecture and Multimedia Gallaecia, 4200-072 Porto, 
Portugal}

\affil[5]{Goldsmiths University of London, Forensic Architecture Research Group, London, SE14 6NW, United Kingdom}
\affil[6]{DEMO Consultants B.V., Department of Research \& Development, Delft, 2628 XJ, Netherlands}
\affil[*]{p.nourian@utwente.nl}
\begin{abstract}
    We present a multi-actor interactive framework for collaboratively forming multi-functional urban building complexes, e.g. in the context of mixed-use housing developments. 
    We propose mechanisms for a mathematical social-choice game that is designed to mediate decision-making processes for city planning, urban area redevelopment, and architectural design (massing) of urban housing complexes, especially to provide mechanisms for the inclusion of less vocal stakeholders.
    The proposed game is effectively a multi-player generative configurator equipped with automated appraisal/scoring mechanisms for revealing the aggregate impact of alternatives; featuring a digital serious gaming approach for participatory design to support transparent and inclusive decision-making processes in spatial design for ensuring an equitable balance of sustainable development goals.
    As such, the game effectively empowers a group of decision-makers to reach a fair consensus by mathematically simulating many rounds of reasonable trade-offs between their decisions, with different levels of interest or control over various types of investments.
    Our proposed gamified design process is general enough to encompass decision-making about the most idiosyncratic aspects of a site related to its heritage status and cultural significance (values and attributes) to the most complex generic aspects such as balancing access to sunlight for the site while respecting ‘the right to sunlight’ of the neighbours of the site, ensuring coherence of the entire configuration with regards to a network of desired closeness ratings, the satisfaction of a programme of requirements, and intricately balancing individual development goals in conjunction with communal goals and environmental design codes. 
    The game is developed fully based on an algebraic computational process on our own digital twinning platform, using open geospatial data and open-source computational tools such as NumPy. 
    The mathematical process consists of a Markovian design machine for balancing design decisions of actors, a massing configurator equipped with Fuzzy Logic and Multi-Criteria Decision Analysis, algebraic graph-theoretical accessibility evaluators, and automated solar-climatic evaluators using geospatial computational geometry. % are "markovian design machines" a well-defined concept? should we add references for that?
\end{abstract}
\begin{document}
% # TODO Add numbered table
% # TODO Add numbered algorithm
% # TODO Add numbered figure
% # TODO Add numbered equations
% # TODO Add page numbers
% # TODO Add line numbers
% # TODO Add appendix pages
% # TODO bring in the narrative flow
% # TODO extract datasets

\flushbottom
\maketitle

\thispagestyle{empty}

%\noindent Please note: Abbreviations should be introduced at the first mention in the main text – no abbreviations lists. Suggested structure of main text (not enforced) is provided below.

\section*{Introduction}
% # TODO elaborate on the problem in the most general sense of it, then connect it to the investment portfolio analogy
The problem framed in this paper generally pertains to the challenge of reaching satisfactory decisions with respect to multiple criteria by a group of actors on allocation of various resources \cite{friedkin_mathematical_2019} into some kind of an investment portfolio, subject to some budget constraints, and quality criteria, with different value systems, possibly conflicting individual goals and various uneven levels of interest and control on the resources and investments. 
While this general picture is recognisable in multiple forms such as organizational decision-making and planning in general, this paper addresses such resource allocation problems in the context of urban architecture and area development, proposing a methodology for facilitating consensus-building and participatory design/decision-making to achieve `consensual satisfaction of multiple criteria'. % # TODO add ref to consensus-building
The generality of such resource-allocation problems on the one hand and the specific challenges arising out of the spatial complexity of the participatory urban-architectural design problem, on the other hand, motivated our mathematical research for devising a game engine for equitable decision-making in the context of spatial developments. % scientific relevance
%#TODO elaborate on the generality to consolidate the scientific relevance
%Additionally, two particular societal challenges motivated the gamification of the participatory design process that extends beyond merely using the mathematical engine of the game, namely dealing with the issue of urban re-development in presence of tangible or intangible cultural heritage as well as diversity, inclusion, and equity in the decision-making process. % societal relevance
%#TODO elaborate with examples to make it more tangible

In short, this research provided an opportunity for testing the potential of games as 'play \& score' mechanisms for facilitating equitable decision-making in dealing with complex urban development problems and their constituent public or community-owned resources. 
See other examples such as \cite{perna_and_2020}, \cite{sousa_planning_2020}, a classification of Simulation Games\cite{djaouti_classifying_2011}, two comprehensive books by Sanoff respectively on participatory planning and design games, \cite{sanoff_community_2000}, \cite{sanoff_design_1978}, a reference on developing games for participatory stakeholder analysis \cite{bots_developing_2003}, a review on city-making games \cite{schouten_games_2017}, an interesting application of games in Transport Planning, explaining the value of the communicative-rational approach to planning in comparison to the technical-rational approach \cite{raghothama_gaming_2015}, the pioneering book of Epstein on Generative Social Science for its analysis of consensus as equilibrium in multi-player games \cite{epstein_generative_2006}, an introduction to Planning Support Systems \cite{geertman_introduction_2015}, a game-theoretical treatise on evolution around equilibrium in multi-player games \cite{webb_population_2007}, a classical book on the virtues of simulation games \cite{raser_raser_1972}, a measure of power in game dynamics similar to our definition of gamification badges \cite{steunenberg_strategic_1999}, a critical overview on the role of optimization models in bench-marking development goals in urban planning \cite{keirstead_changing_2013}, a gamified participatory design/planning framework \cite{slingerland_together_2020}, and a succinct overview of complexity in urban planning\cite{crawford_what_2016}. %#TODO weave the references to the argumentation for clearifying the gap and scope of this paper 
Correspondingly, the purpose of the proposed game is to provide a non-reductionist model for decision-support in complex decision-making problems concerned with spatial design with constrained resources and a multitude of model sustainability goals (see another game concerned with multi-actor sustainable development in Monechi et al. \cite{monechi_finding_2021}). 
The specific sustainable development goals are defined as instances of three archetypal categories, for each of which we have considered a model example:
\begin{itemize}
    \item social-economic equity: w.r.t. fairly distributing costs and benefits of a development. This objective is ensured already by proposing a participatory opinion pooling mechanism (the term opinion pool dates back to Stone \cite{stone_opinion_1961} \& De Groot \cite{degroot_reaching_1974}); 
    \item economic-environmental efficiency: w.r.t. scoring the change of allocation per site, and the degree to which the massing distribution blocks the solar potential of the neighbourhood;
    \item environmental-social comfort: w.r.t. stated preferences (closeness ratings) between the compartments of the district as well as daylight potential of the district. 
\end{itemize}
The proposed game was conceived as a modular and scalable platform that could incorporate various types of evaluation mechanisms on a Digital Twin of an urban district for prototyping Spatial Decision-Support Systems. 

A key factor in forming the proposed gamified 'social choice mechanism' (q.v. Jackson's definition \cite{jackson_mechanism_2014} \& a similar recent formulation \cite{bai_decision-making_2020}) is the subtle difference between optimisation and gamification (regulated group decision-making with scoring mechanisms) approaches to such policy, planning, and design problems involving resource allocation. 
As stated by Bots \& Herman \cite{bots_developing_2003}, if stakeholders know each other’s controls and interests and if they agree on fixing some average of these interests and weights of criteria, then the negotiation might be modelled as a puzzle for which some optimal solution can be found. %#TODO refs from game theory on efficiency properties, fairness properties, strategic properties, and participation properties are reuired here. start from: https://en.wikipedia.org/wiki/Fractional_approval_voting#cite_note-:4-5
But if there are uncertainties in the definition of the problem and different views towards the objectives, especially if there are power differences, then stakeholders (hereinafter referred to interchangeably as agents or actors) will play strategic ‘games’ that may produce complicated outcomes, better or worse than ideally would be possible depending on whether they would be cooperative or overly competitive. 
%In other words, the game-theoretical dimension of the problem emerges naturally out of the non-trivial power-balances and uncertainties concerning the ultimately unpredictable changes of agendas of players. % Shervin: this is quite relevant, why is it commented?
To this end, we propose to measure some game-theoretical indicators of cooperation (contribution to the common objective) and competition (sagacity for achieving one's ends without much means) to positively reinforce constructive negotiations during the game (vide infra, \ref{fig:game_scores}).
%#TODO separate the background information into background section

%If the quantification of goals and constraints (which might be as important or even more important than the goals per se) is subject to uncertainty and multiple value systems of multiple actors, one might question the utility or reliability of an optimization process for reaching a satisfactory solution as compared to a conventional or regulated group decision-making process based on negotiations. 
As has been argued by the Nobel Laureate Herbert Alexander Simon \cite{simon_sciences_1996}, it is common-knowledge that one can only refer to `the optimal' in presence of a single objective. In the so-called multi-objective optimization problems, the reality is that whether we use games, heuristics, meta-heuristics or even mathematical programming methods from Operations Research, we can only 'satisfy' the problem as formulated in presence of various simplifications, abstractions, and approximations, but 'the optimal solution' does not exist as such. 
Thus, especially if the decision-outcome is to be accepted by a group of human actors, playing a purposeful game or going through a structured negotiation process can be arguably more relevant and effective than attempting to reduce a multi-actor multi-criteria decision making problem into a multi-objective optimization problem. 

It is noteworthy that the ideas of utilizing simulation games for understanding decision-making processes and game theoretical approaches to city planning date back at least to the 1970's \cite{raser_raser_1972}. 
Susan Batty clearly describes the kind of complexity arising out of uncertainties concerning the decisions, agendas/value-systems, and the costs/benefits of one actor's decisions for oneself and the influence of the decisions of other actors on one's interests \cite{batty_game-theoretic_1977}. 

Michael Batty's formative works on this subject area reveal how the complexities arising out of the spatial context of the problems make the resource allocation problems more challenging and mathematically interesting at the same time \cite{batty_-new-science--cities_2013, batty_cities_2009}.
In  "Evolving a Plan" \cite{batty_evolving_2016} Batty proposes a process of Opinion Pooling dating back to French 1956 \cite{french_jr_formal_1956}, and Harary 1959 \cite{harary_criterion_1959}, based on his earlier idea of Markovian Design Machines. 
The typical problem addressed in that book chapter is a recurrent theme in Batty's work pertaining to the human complexity of multi-actor (multi-agent) decision making and finding a satisfactory plan of actions (resource allocation) with respect to a set of objects (factors in his formulation). 
The readers interested in the mathematical analysis of opinion dynamics are referred to the work of Jia et al \cite{jia_opinion_2015}. %# this sentence is educational, maybe we can use this reference in argumentation in a more integrated way
The basic set up of the problem scrutinized by Batty is essentially a problem of resource allocation to multiple objects/sites of interest in an urban redevelopment setting. However, without loss of generality, a similar process can be applied to non-spatial problems of planning accordingly.
In fact, what is presented here can be thought of as an extension and a generalization of the work of Batty on Markovian Design Machines and their application to spatial design problems. 
Our extended problem formulation, in particular, considers that there might be different colours (sorts) of resources to be allocated to a particular target site (could also be a part of a portfolio or any such object of interest). 

\section*{Proposed Framework}
Here we present our proposed framework for structuring such generic problems in participatory spatial configuration problems.
\subsection*{Problem Statement}
The game is to facilitate participation in decision making for three types of design/planning problems dubbed as pre-planning, planning, and massing problems. 
%Additionally, an illustrative zoning solution provider is added to the game; but the problem of zoning \emph{per se} falls out of the scope of the paper. 
Note that the generic design problems dubbed here as massing and zoning are also known by other names such as configuration problems (q.v. an influential framework by Yona Friedman \cite{friedman_toward_1975} and a congruent definition in a generative design framework \cite{azadi_godesign_2021}). 
Although it is important to state the preconditions and assumptions underlying the problem definitions, in the interest of generality, the paper directly goes into the most abstract definition of each problem, for a visual summary see Figure \ref{fig:problem} and see the section Problem-Specific Settings pursuant to the three identified problems.

\subsubsection*{A) The Pre-Planning Problem}
This problem concerns the collation of stated preferences from a group of actors, with the objective to build a consensus by simulating a negotiation and averaging process through a Markov-Chain (Opinion Pooling). 
The problem is defined as determining the amount of 'investment' in each category of investment objects (hereinafter referred to as colours), which in this case refer to the distinct types of spaces designated for accommodating different activities (e.g. residential, commercial, cultural, or other sorts of spaces, this is known as a Programme of Requirements in the design and planning jargon). 
Given a set of such colours, and a set of portfolios (sites or buildings in this case), the actors are to decide how much of each type of investment must be made in each portfolio (how much of each type of space in each site/building). 
What distinguishes this from a trivial problem of averaging the votes is that the actors have various degrees of interest and control over these various types of investments, denoted respectively by the three-dimensional matrices $\mathbf{X}_{m\times n \times o}:=[X_{i,j,k}]_{m\times n \times o}$ and $\mathbf{C}_{n\times m \times o}:=[C_{j,i,k}]_{n\times m \times o}$, where $m$, $n$, and $o$ respectively denote the number of actors, sites, and colours. 
The solution to this problem is supposed to be a plan consisting of the amount of desired lettable/saleable net floor space of each colour type per each site, practically a matrix $\mathbf{A}_{n \times o}^{(t)}:=[A_{j,k}]_{n \times o}^{(t)}$, where $n$ denotes the number sites and $o$ denotes the number of colours, and the superscript $(t)$ denotes a time stamp referring to the discrete time of the game, colloquially referred to as a round of playing.

\subsubsection*{B) The Planning Problem} 
The planning problem concerns finding the exact amount of volume of each type/colour of space in each site, as closely as possible to the decided amount of net (lettable) coloured areas, matching the prescribed totals for each colour on the entire site, and scaled aptly to integer volumetric quanta in order to yield the expected amounts of net lettable floor area per colour (apropos the consensual decisions of the actors).
%It is desired to find the volume of each colour per site as close as possible to a given distribution indicating the local tendency of each site for having certain area of each colour such that the total amounts of allocated colours per site add up exactly to the total amounts of required colour areas for the entire district. 
Additionally, a given set of scaling factors indicate the volume required per each type of coloured space for realizing the desired net floor areas per site, and so, from this point forward the distributions concern the volumetric spaces rather than surface areas. 
 
Whilst the exact values of the latter scale factors are irrelevant to the subject matter of the paper, their practical existence and difference in terms of gross volume to net lettable areas is undeniable and thus typical values are considered to ensure the generality of the formulation. 
These scale factors, however, are treated as the "advanced settings" of the game and set by the game master rather than the participant players. 
The adjustment of the local mixes of colours to the expected global mix ratios, dubbed as the District Level Program, denoting Required Surface-Area per Color $\mathbf{y} := [y_{k}]_{o \times 1}$, is performed by means of the Iterative Proportional Fitting. %#TODO ref to pirouz's paper on proportional fitting
Two additional variable bounds are to be respected in this problem, namely the maximum building height and maximum gross floor area per each site. 
The solution to this problem is a plan indicating the integer number of volumetric cells of space to be built of each colour in each site; this is formally a matrix $\mathbf{V}_{n \times o}^{(t)}:=[V_{j,k}]_{n \times o}^{(t)}$.

\subsubsection*{C) The Massing Problem}
\begin{figure}[p]
\centering
\includegraphics[width=16.5 cm]{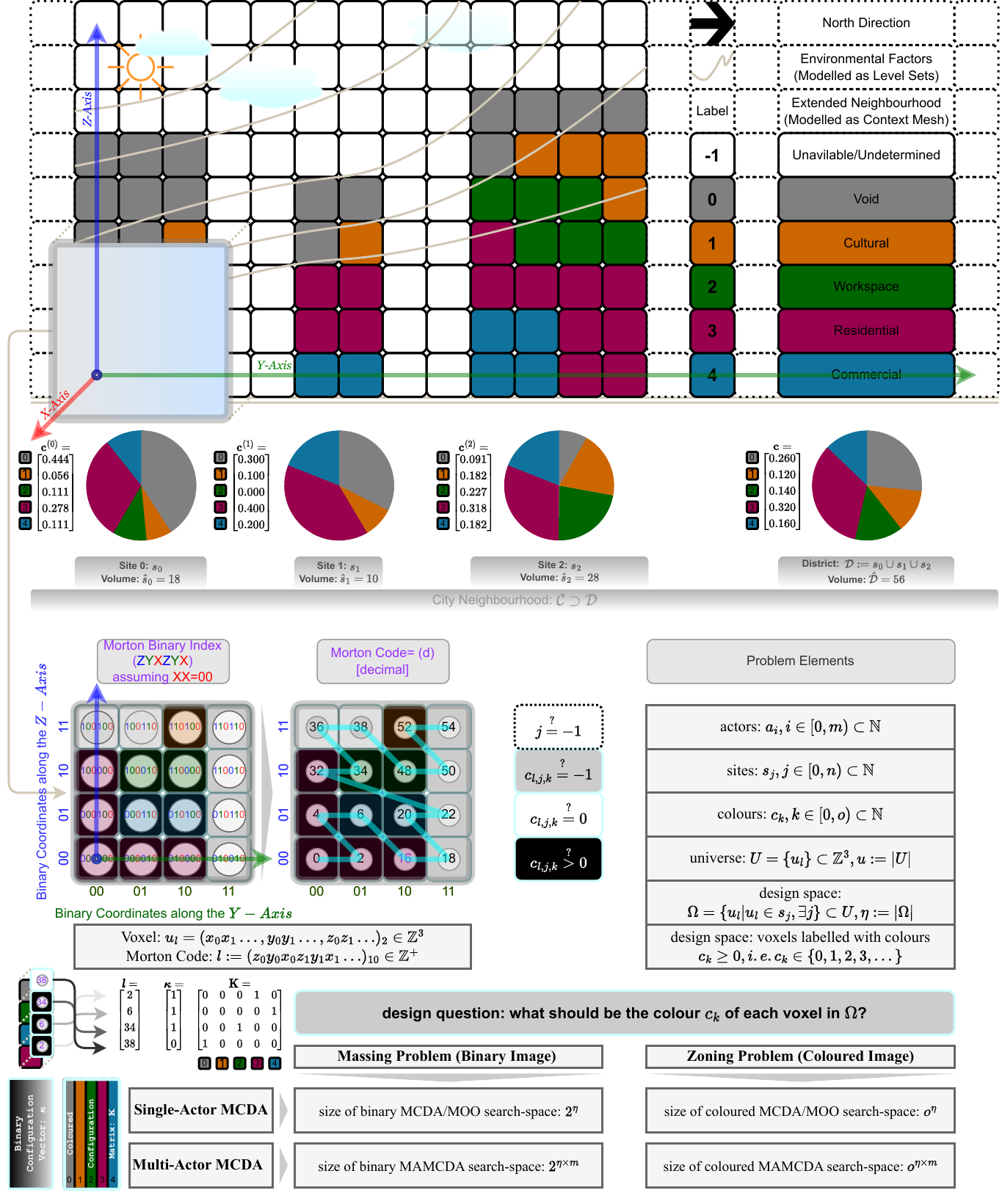}
\caption{Illustrated Formulation of a Toy-Problem}
\label{fig:problem}
\end{figure}

Given the total quanta of the volumetric coloured spatial units per site, the polygonal surface geometry of each site, a regular grid of volumetric spatial units (volumetric pixels or voxels), stated weights of importance of a number of massing quality criteria, procedures for ex-ante assessment of the said quality criteria, a '3D context mesh' (a discrete surface consisted of vertices and faces commonly known as a 3D city model, denoted as $\mathcal{M}=(V,F)|V\subset \mathbb{R}^3$) as to which some costs are to be computed, the problem is to determine the shape of a voxelated volume, i.e. the index of the coloured voxels, per each voxelated site so as to satisfy the quality criteria as optimally as possible. 
Even though the coloured version of this problem (hereinafter referred to as zoning) goes far beyond the scope of this paper, in the latest implementations of the game we have added a simplistic zoning procedure that determines the exact location of each colour within each site for better illustration of the outcome.
However, the exact placement of colours within each site, as long as the total amount of coloured space has the same outer envelope shape, does not affect the aggregate quality criteria reported as objectives ($\mathbf{q}$, vide infra \ref{sec:magma}), and so, the massing problem is one level of abstraction higher than the zoning problem. The solution to the massing problem is in the form of the Mutually Exclusive and Collectively Exhaustive (MECE) indices of coloured voxels in the discrete image of the district, considering a dummy-undefined colour for the uncoloured (undesignated) spaces (see Figure \ref{fig:problem}). 

\subsubsection*{Problem-Specific Settings}
\label{sec:problemsettings}
%Participatory decision making is a process that is founded on the negotiations of different actors regarding their preferences for the subject matter. 
%In the context of urban design, the complexity of the subject matter occludes the negotiation as the actors are not able to easily appraise alternative decisions and extrapolate the consequences of such decisions. 
%Therefore, the communication of concerns and preferences becomes extremely hard for actors and negotiations results in unnecessary compromises or even non-consensual conclusions.

An urban district with a mix of already existing uses is to be densified to accommodate a larger group of incoming inhabitants. 
A group of actors are to decide on the colours (i.e. the programmatic allocation of labels) of a number of sites comprising a larger district. 
Each colour has a different lettable/useable area to volume ratio, due to average space height, different needs for corridor space, and alike.The ratio of Lettable Floor Space (LFS) to Gross Floor Area (GFA) is different and given per each colour (we regard such adjustments as advanced settings).
The cost of change (implied retrofitting, demolition, or otherwise) is also different and assumed per site and color (another set of items from advanced settings). Additionally, each site has potentially different design codes that can be translated to maximum allowed height or a build-able volumetric envelope. 
The actors have various levels of control and interest (varied per color) over the district, specified per parcel of land. 

The actors can specify their intent (proposal) for the ideal mix of uses (colours) to be allocated to each site. 
They can specify the desired levels using levels ranging between 0 and 1, which are relativised as portions of colours or coloured surface areas. 
Similarly the control of multiple actors over each site has to add up to 100\%. Formally, these two conditions translate to the Interest and Control matrices being row-stochastic, which are ensured in the back-end of the system.

The 3D design space is discretised into a regular grid of volumetric pixels (voxels) indexed with globally unique addresses obtained as their Morton Codes (see Figure \ref{fig:problem}).

\subsection*{Problem Formulation}
In the following we shall compare two alternative formulations of the problem and show how the gamified negotiation problem differs from the optimization problem formulation (cf. Illustrations in the supplementary materials and the Figure \ref{fig:problem} for complexity analysis w.r.t. problem size in MOO and MCDM settings).

\subsubsection*{As an Optimisation Problem}
Given a set of locations with known areas, a regularly discretized 3D spatial domain, a programme of requirements for the district consisting of the said locations specifying the amount of surface area per programmatic label (colour) and the maximum amount of built area per site, it is desired to find the most satisfactory programmatic allocation for the entire district subject to the following constraints and optimality criteria (if weighted similarly by all actors involved): 

\begin{itemize}
    \item Constraints
    \begin{itemize}
        \item The total allocated area per color must be the same as the given district-level programme
        \item The maximum allocated area per site must not exceed the maximum allowed Floor Space Index per site
    \end{itemize}
    \item Objectives (illustrative)
    \begin{itemize}
        \item Maximizing the visibility of the sun, the sky, a landscape object of choice (for the district and its neighbourhood within a radius) (figuratively introduced as goals related to Planet \& People).
        \item Maximizing the similarity of the closeness rates after allocation with the initial given closeness ratings in a REL chart (figuratively introduced as goals related to People \& Prosperity).
        \item Minimizing change of area allocation per site (figuratively introduced as goals related to Prosperity \& Planet).
    \end{itemize}
\end{itemize}

\subsubsection*{As a Gamified Negotiation Problem}
Even though the single-actor optimization problem is not directly addressed in the paper, it is necessary to consider it as a baseline for understanding the size of the problem. It is straightforward to see that the size of the search-space, i.e. the complexity of finding a solution (a configuration satisfying the constraints and objectives) for the problem by mere chance, e.g. for a 'monkey behind a type-writer', corresponds to the number of possible configurations in the discretised domain, that is $2^\eta$ for the massing (black \& white colouring) problem.
However, there are also agendas and different value systems complicating the problem, thus the size of the search space rises to $2^{\eta \times m}$ (for $m$ nitpicking participants without the game mechanisms, that is, see Figure \ref{fig:problem}).

In order to positively reinforce constructive negotiation activities two badges are defined to be issued in each round of the game to the most cooperative player and the most competitive contributor based on a definition of Power Surplus (Dearth) that goes beyond the simplistic definition of winners and losers based on closeness of the final decision to the decision of the actor.
In the gamified negotiation problem, there exist $m$ opinions on the $n\times o$ distribution of colours on sites. 
After the stages of Opinion Pooling and Iterative Proportional Fitting (explained further), there will be one such distribution of coloured volumes on sites that can be compared to the expressed opinions of each one of the actors to determine the gainer, the player and the contributor of the round as follows (details further explained in the supplementary materials):

The "Gainer of the Round" is the actor with the most similar interest matrix to
the collective decision, formally:
\begin{equation}
    \arg\min_{i} \|\mathbf{X}^{(t)}[i,:,:]-\mathbf{A}\|_{F}.
\end{equation}

The  badge of honour "Player of the Round" is defined as
the actor with the most similar pattern of `the negative parts of their Power Surplus matrix' (dubbed $\mathbf{\pi}_{\ominus}^{(t)}$) to
the collective decision, or formally as:
\begin{equation}
\arg\min_{i} \left[\pi_{\ominus}^{(t)}[i]\right]_{m\times1}
\end{equation}

The badge of honour "Contributor of the Round" is defined as
the actor with the most similar pattern of `the positive parts of their Power Surplus matrix' (dubbed $\mathbf{\pi}_{\oplus}^{(t)}$) to
the collective decision, or formally as:
\begin{equation}
\arg\min_{i} \left[\pi_{\oplus}^{(t)}[i]\right]_{m\times1}
\end{equation}

While the derivation of the gainer/looser badges is trivial, the player/contributor badges have been derived through a complex process illustrated in the Figure \ref*{fig:badges}. See the supplementary information document for the derivation process.
\begin{figure}[H]
    \centering
    \includegraphics[width=17 cm]{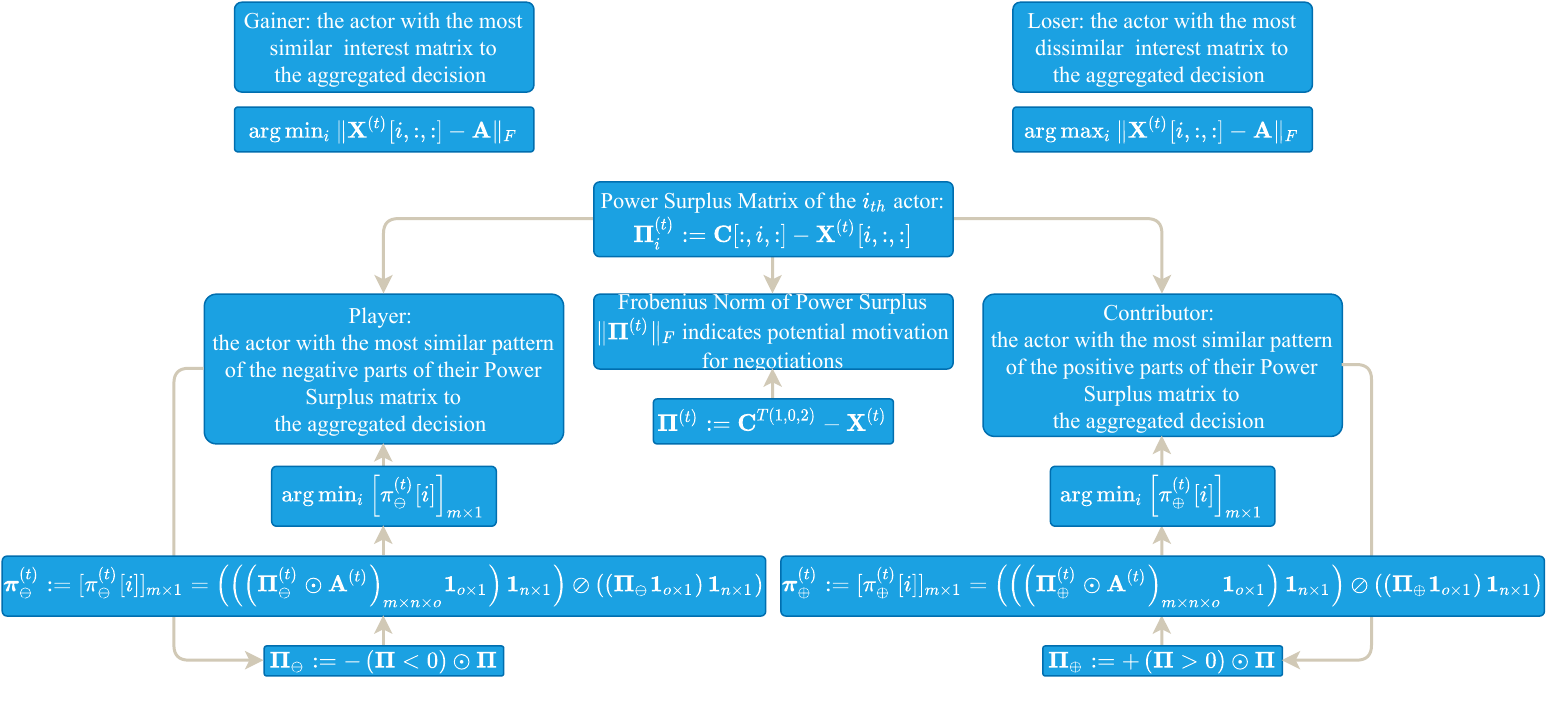}
    \caption{Flowchart indicating the derivation of the Gainer (top-left), Player, and Contributor badges of the game. The Loser badge is not communicated.}
    \label{fig:badges}
\end{figure}

\begin{figure}[ht]
\centering
\includegraphics[width=15 cm]{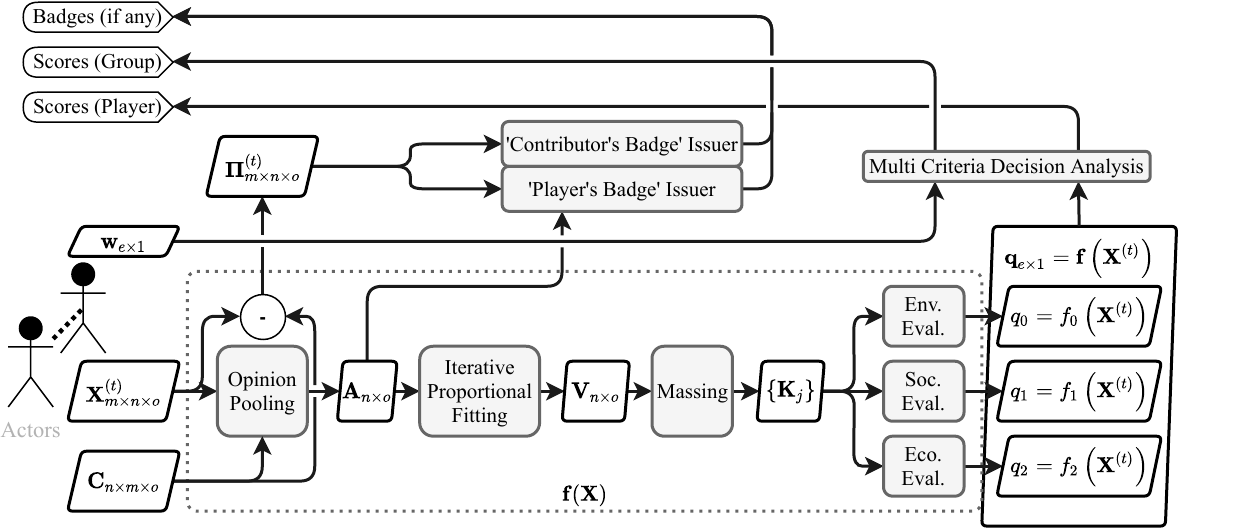}
\caption{Illustrated formulation of the problem as a gamified negotiation problem, cf. a more detailed schema and an alternative formulation as a Multi-Objective Optimization in the Supplementary Information}
\label{fig:gamification-problem}
\end{figure}

In this formulation, we propose to firstly build a fair consensus between the actors and motivate them to engage in negotiations by issuing the badges introduced above and inform them in the meantime on the environmental, economic, and social quality of the allocation decision emerging out of the expressed opinions (weighted individual and group scores). The game engine features the following generative process in each round of voting (see the schematic data-flow diagram in the supplementary materials for a visual overview). 

\section*{Proposed Methods}
Here we introduce the proposed processes for participatory design of a city neighbourhood, from the basic discretization of design for converting it into a Multi-Criteria Decision-Making (MCDM) problem, to the creation of a consensus-building procedure (Opinion Pooling), the adjustment of the portfolio according to a [hypothetically] prescribed master-plan of the area by a municipal authority (Iterative Proportional Fitting), generative massing based on a Fuzzy Multi-Criteria Decision Analysis, a very basic (illustrative) zoning process formulated as a clustering problem, and the gamification of the process by encouraging strategic thinking through issuance of mathematically defined badges dubbed: the gainer, the contributor, and the player of the round.

\subsection*{Gamified Generative Design}
The following processes are proposed to solve the three problems identified in the framework: A) Pre-Planning, B) Planning, and C) Massing. These processes are generative in that together they generate more information content than they receive from the players and they are looped through a gamified cycle illustrated in Figure \ref{fig:gamification-problem} hence the term gamified generative design.
\subsubsection*{A) Algebraic Opinion Pooling}
The purpose of this process is to simulate the convergence of a fair/equitable pooling of opinions of multiple actors on a distribution of coloured resources over sites, and produce the matrix $\mathbf{A}^{(t)}$. Suppose a group of actors are to decide on some degree of action (an amount of investment) on a group of [portfolio] objects (sites in our terminology). 
Batty  \cite{batty_evolving_2016} states that if the actors have all the same degrees of interest and control over the objects/factors in question, then the problem is trivial and simple. 
However, if the degrees of interest and control are not the same then the problem can be formulated as finding a unanimously agreeable or fair consensus amongst the agents with respect to the amount of change so as to deviate minimally from all expressed opinions based on perceptions of actors on their stakes. 
In other words, an actor might have a high degree of interest in a factor but little control over it and vice versa. These differences between interests and control levels define the non-trivial bipartite relative interaction networks:
\begin{itemize}
    \item interactions between actors (agents) over sites (factors); this is dubbed as 'the primal problem'
    \item interactions between sites (factors) over actors (agents); this is dubbed as 'the dual problem'
\end{itemize}
%In our formulation, we also consider that the actors may have different sorts of interests as per different types of investments per portfolio objects (sites). 
Thee different types of interests/investments are emblematically referred to as colours in our formulation. 
Effectively, we generalize the process by iteratively solving the opinion-pooling problem for each colour.
Table \ref{tab:opinion-pooling} summarises the derivation of the generalized algebraic opinion-pooling method, and Algorithm \ref{alg:opinion-pooling} shows it in a reproducible and scalable form.

\begin{table}[ht]
\centering
\resizebox{\columnwidth}{!}{%

\begin{tabular}{l|c||c}
\toprule
Problem Type & 
\textbf{Primal Problem} &
\textbf{Dual Problem}
\\ \midrule
Description  & 
\makecell{distribute an investment [colour] amongst actors\\ through an interaction network across sites} & 
\makecell{distribute an investment [colour] amongst sites\\ through an interaction network across actors} \\
Network      & 
$\mathbf{P}=\mathbf{X}\mathbf{C}$ & 
$\mathbf{Q}=\mathbf{C}\mathbf{X}$ \\
Markov Chain      & 
$\boldsymbol{\alpha}^{(t)}={\alpha}^{(t-1)}\mathbf{P}$ & 
$\boldsymbol{\beta}^{(t)}={\beta}^{(t-1)}\mathbf{Q}$  \\\midrule
\makecell[l]{Steady State\\ (definition)}      & 
\makecell{\(\displaystyle \boldsymbol{\alpha}:=\lim_{t \rightarrow \infty}\boldsymbol{\alpha}^{(t)}\)\\
\(\boldsymbol{\alpha}\mathbf{P}=\boldsymbol{\alpha}\)\\  \(\boldsymbol{\alpha}\mathbf{1}=1\)}& 
\makecell{\(\displaystyle \boldsymbol{\beta}:=\lim_{t \rightarrow \infty}\boldsymbol{\beta}^{(t)}\)\\
\(\boldsymbol{\beta}\mathbf{Q}=\boldsymbol{\beta}\)\\ \(\boldsymbol{\beta}\mathbf{1}=1\)}\\\midrule
\makecell[l]{Steady State\\ (solution\cite[~pp.250-252]{nourian_configraphics_2016})}      & 
\makecell{\(\displaystyle \boldsymbol{\alpha}\left [ \left (\mathbf{I}_{m\times m}-\mathbf{P}\right )|\mathbf{1}_{m\times 1}\right]=\left [\mathbf{0}_{1\times m}|1 \right]\) \\\\
\(\displaystyle \underbrace{\left[ \left(\mathbf{I}_{m\times m}-\mathbf{P}\right )|\mathbf{1}_{m\times 1}\right]^T)}_\text{$\mathbf{M}$} \underbrace{\boldsymbol{\alpha}^T}_\text{$\mathbf{x}$}=\underbrace{\left [\mathbf{0}_{1\times m}|1 \right ]^T}_\text{$\mathbf{a}$}\)\\\\
\(\displaystyle\boldsymbol{\alpha}^T=\arg\min_{\mathbf{x}}{\|\mathbf{M} \mathbf{x}-\mathbf{a}\|^2_2}\)}& 
\makecell{\(\displaystyle \boldsymbol{\beta}\left [ \left (\mathbf{I}_{n\times n}-\mathbf{Q}\right )|\mathbf{1}_{n\times 1}\right]=\left [\mathbf{0}_{1\times n}|1 \right]\) \\\\
\(\displaystyle \underbrace{\left[ \left(\mathbf{I}_{n\times n}-\mathbf{Q}\right )|\mathbf{1}_{n\times 1}\right]^T)}_\text{$\mathbf{N}$} \underbrace{\boldsymbol{\beta}^T}_\text{$\mathbf{y}$}=\underbrace{\left [\mathbf{0}_{1\times n}|1 \right ]^T}_\text{$\mathbf{b}$}\)\\\\
\(\displaystyle\boldsymbol{\beta}^T=\arg\min_{\mathbf{y}}{\|\mathbf{N} \mathbf{y}-\mathbf{b}\|^2_2}\)}\\
Duality      & 
$\boldsymbol{\alpha}=\boldsymbol{\beta}\mathbf{C}$ & 
$\boldsymbol{\beta}=\boldsymbol{\alpha}\mathbf{X}$ \\
\bottomrule
\end{tabular}%
}
\caption{\label{tab:opinion-pooling}A summary of the opinion pooling process}
\end{table}

\begin{algorithm}[htb]
    \caption{Algebraic Opinion Pooling (Vectorised and Generalised to Categorical Investments)}\label{alg:opinion-pooling}
    \KwData{$\mathbf{X}^{(t)}_{m\times n\times o}, \mathbf{C}_{n\times m\times o}$}
    \KwResult{$\mathbf{B}_{n\times o}:= [\boldsymbol{\beta}^T_0|\boldsymbol{\beta}^T_1|\dots |\boldsymbol{\beta}^T_k|\dots ]$}
    \For{$k \in [0,o)$}{
    $\mathbf{X}_k\gets \mathbf{X}[:,:,k]$\;
    $\mathbf{C}_k\gets \mathbf{C}[:,:,k]$\;
    $\boldsymbol{\mYa}\gets \text{diag}(\mathbf{X}_k\mathbf{1}_{n,1})$\;
    $\mathbf{R}\gets \text{diag}(\mathbf{C}_k\mathbf{1}_{m,1})$\;
    $\mathbf{X}_k\gets \boldsymbol{\mYa}^{-1}\mathbf{X}_k$\Comment*[r]{Ensure that $\mathbf{X}$ is row stochastic}
    $\mathbf{C}_k\gets \mathbf{R}^{-1}\mathbf{C}_k$\Comment*[r]{Ensure that $\mathbf{C}$ is row stochastic}
    $\mathbf{Q} \gets \mathbf{C}\mathbf{X}$\;
    $\mathbf{N} \gets \left[\mathbf{I}_{n \times n}-\mathbf{Q}|\mathbf{1}_{n\times 1} \right]^T$\;
    $\mathbf{b} \gets \left[\mathbf{0}_{1\times n}|1\right]^T$\;
    $\mathbf{y}\gets \mathbf{0}_{n\times 1}$\Comment*[r]{Initialization}
    $\displaystyle \boldsymbol{\beta}^{T}_k\gets \arg\min_{\mathbf{y}}{\|\mathbf{N} \mathbf{y}-\mathbf{b}\|^2_{2}}$\Comment*[r]{Least-Squares Solution}
    $\mathbf{B}[:,k]\gets \boldsymbol{\beta}^T$\Comment*[r]{The allocation matrix dubbed $\mathbf{A}$ outside this algorithm}
      %\eIf{$N$ is even}{
      %  $X \gets X \times X$\;
      %  $N \gets \frac{N}{2}$ \Comment*[r]{This is a comment}
      %}{\If{$N$ is odd}{
      %    $y \gets y \times X$\;
      %    $N \gets N - 1$\;
      %  }
      }
    \end{algorithm}

\subsubsection*{B) Algebraic Iterative Proportional Fitting}
The purpose of this process is to arrive at a matrix $\mathbf{V}^{(t)}$ representing the colour volumes for each site with row sums and column sums equalized respectively to the district-level Programme of Requirements and the volume capacities of the sites, while remaining as close as possible to the tensor direction of $\mathbf{A}^{(t)}$. A generic problem that occurs in planning with the abstractions presented here is that there might be a prescribed or desired district level programme consisting of a given distribution of the categorical (coloured) investments, while each site (portfolio object) that is being invested in has some capacity for containing coloured amounts of investments, it may locally prefer to have more or less of some amounts. In other words, the players effectively aim for certain local distributions of colours without being able to tediously ensure that the local distributions add up to the two global distributions: a distribution of coloured area amounts for the whole district, and a distribution of the site capacities \cite{nourian_interactive_2013}. This mathematically corresponds to the formation of a ``contingency table'' or (in the special case that the marginal totals are stochastic themselves) a doubly-stochastic matrix as closely as possible to a [possibly row/column stochastic] desired distribution.
The Iterative Proportional Fitting \cite{deming_least_1940}\cite{stephan_iterative_1942}\cite{hunsinger_iterative_2008} procedure is meant to adjust the entries of the allocation matrix $\mathbf{A}=[A_{j,k}]_{n\times n}$, while keeping it similar to the original matrix, such that its row sums and column sums reach the predefined capacity distributions. Algorithm \ref{alg:iterative-proportional-fitting} presents our reproducible algebraic method for scalable proportional fitting.

\begin{algorithm}[H]
\caption{Algebraic Iterative Proportional Fitting)}\label{alg:iterative-proportional-fitting}
\KwData{$\mathbf{A}_{n\times o}$,  target row sums: $\boldsymbol{\mathfrak{r}}$, target column sums:$\boldsymbol{\mathfrak{c}}$, square error tolerance $\varepsilon$, maximum iterations $\mathfrak{t}$}\Comment*[r]{Ensure that row-sums and col-sums both add up to the same total amount}
\KwResult{$\Bar{\mathbf{A}}_{n\times o}: \Bar{\mathbf{A}}_{n\times o}\mathbf{1}_{o\times 1}=\mathbf{1}_{n\times 1} \; \&\; \Bar{\mathbf{A}}^T_{o\times n}\mathbf{1}_{n\times 1}=\mathbf{1}_{o\times 1}$}
$\boldsymbol{\mathfrak{R}}\gets \text{diag}(\boldsymbol{\mathfrak{r}})$\;
$\boldsymbol{\mathfrak{C}}\gets \text{diag}(\boldsymbol{\mathfrak{c}})$\;
$\boldsymbol{\rho}_{n\times 1}\gets\mathbf{A}_{n\times o}\mathbf{1}_{o\times 1}$\Comment*[r]{Initial row-sums}
$\boldsymbol{\kappa}_{o\times 1}\gets\mathbf{A}_{o \times n}^T \mathbf{1}_{n\times 1}$\Comment*[r]{Initial col-sums}
$t\gets 0$\;
$e\gets (\boldsymbol{\rho}-\boldsymbol{\mathfrak{r}})^T(\rho-\mathfrak{r})+(\boldsymbol{\kappa}-\boldsymbol{\mathfrak{c}})^T(\boldsymbol{\kappa}-\boldsymbol{\mathfrak{c}})$\;
\While{$\displaystyle((e> \varepsilon)\wedge(t< \mathfrak{t}))$}{
%$\boldsymbol{\rho}_{n\times 1}\gets\mathbf{A}_{n\times o}\mathbf{1}_{o\times 1}$\;%building current row-sums
%$\boldsymbol{\kappa}_{o\times 1}\gets\mathbf{A}_{o \times n}^T \mathbf{1}_{n\times 1}$\;%building current col-sums
$\mathbf{P}\gets \text{diag}(\mathbf{A}\mathbf{1})$\;
$\mathbf{A}\gets \boldsymbol{\mathfrak{R}}\mathbf{P}^{-1}\mathbf{A}$\;
$\mathbf{K}\gets \text{diag}(\mathbf{A}^T\mathbf{1})$\;
$\mathbf{A}\gets \mathbf{A}\mathbf{K}^{-1}\boldsymbol{\mathfrak{C}}$\;
$\boldsymbol{\rho}\gets\mathbf{A}\mathbf{1}$\Comment*[r]{Current row-sums}
$\boldsymbol{\kappa}\gets\mathbf{A}^T \mathbf{1}$\Comment*[r]{Current col-sums}
$e\gets (\rho-\mathfrak{r})^T(\rho-\mathfrak{r})+(\kappa-\mathfrak{c})^T(\kappa-\mathfrak{c})$\Comment*[r]{Total squared error}
$t\gets t+1$\;
}
%https://shantoroy.com/latex/how-to-write-algorithm-in-latex/
%\SetKwFunction{FSub}{IPF}
%\DontPrintSemicolon
%\SetKwProg{Fn}{Def}{:}{}
%  \Fn{\FSub{$first$, $second$}}{
%        a = first\;
%        b = second\;
%        sum = first - second\;
%        \KwRet sum\;
%  }\; 
\end{algorithm}
\subsubsection*{C) MAGMA: Multi-Attribute Gradient-Driven Mass Aggregation}
\label{sec:magma}
The purpose of the massing process is to allocate the spaces according to the amounts given in $\mathbf{V}^{(t)}$, arrive at a mass-configuration dubbed $\mathbf{K}$ (which is only an arbitrarily coloured version of the binary mass-configuration vector $\boldsymbol{\kappa}$)  focusing on achieving the highest total (multi-criteria) value for the allocated spaces, w.r.t. the quality-criteria or outcomes of interest given as $\{q_\iota\}$. 
The method proposed here is based on computing the sensitivities of some aggregate outcomes of interest to the existence or absence of each discrete cellular decision variable (voxel $u_l\in\Omega\subset U\subset \mathbb{Z}^3$) that is indexed with a globally unique index (Morton Index). We dub these fields of sensitivities $\boldsymbol{\varphi}^{(\iota)}:=[\varphi_{l}^{(\iota)}]_{\eta\times 1}$. 

The major breakthrough of the generative design process is that it utilizes the sensitivities (gradient) of the aggregate evaluation scores (a.k.a. Key Performance Indicators, often abbreviated as KPIs) to decide on the existence/absence of each volumetric cellular space domain (i.e. their density or opacity) at a disaggregated spatial level. 
Thus, for each volumetric cell we aim to know the contribution of the cell in question to all outcomes of interest. 
In other words, these two sorts of spatially aggregate and spatially disaggregated evaluation scores are mathematical duals in that the former is the integral of the latter and the latter is (can be considered to be) the gradient of the former. %The process is expressed mathematically in the supplemental materials.

%The generative process works by deducing the total value of each cell w.r.t. all outcomes of interest aggregated using a Fuzzy Logic T-Norm aggregator (Paraboloid AND Aggregator) as introduced in \cite[p.202]{nourian_configraphics_2016}.

%These scalar fields of sensitives are effectively the gradients of the integral outcomes of interest. 
%in other words, the aggregate outcomes of interest shown in the multi-criteria gamified negotiation and multi-objective design optimization processes are only the integral versions of the quality criteria, which have been aggregated (integrated) spatially. 
Thus the following equation shows that the aggregate (integral) outcomes of interest ($\mathbf{q}:=[q_{\iota}]_{e\times 1}$) are obtained simply by integrating the gradients (sensitivities):
\begin{equation}
    \boldsymbol{\varPhi}:=[\varphi_{l}^{(\iota)}]_{\eta\times e},
\end{equation}
meaning that if the desegregated sensitivities are known the aggregate KPI can be easily derived as discrete integrals in the form of:
\begin{equation}
    \mathbf{q}={1}_{\eta \times 1}^{T}\boldsymbol{\varphi}\mathbf.
\end{equation}
The reverse process, i.e. the derivation of the disaggregated evaluation scores however, is often much more complex and domain-specific and thus out of the scope of this paper.

The massing process then forms a multi-criteria total value ($\upsilon_l$) for each cell based on a Fuzzy Paraboloid AND Aggregation (a T-Norm function) introduced in \cite[202]{nourian_configraphics_2016} of the disaggregated quality-criteria pertaining to the location in question:
\begin{equation}
    \upsilon_{l}:=\bigcap_{\iota}\varphi_{l,\iota}=\prod_{\iota}\varphi_{l,\iota}^{w_\iota},
\end{equation}
 
where the eventual weights ($w_\iota$) of quality criteria are the averages of weights submitted by all actors ($\mathbf{W}:=[w_{\iota,i}]_{e\times m}$), i.e. 
\begin{equation}
    \mathbf{w}:=[w_\iota]=\frac{1}{m}\mathbf{W}\mathbf{1}_{m\times 1}
\end{equation}
Then the massing process picks the top number of total required voxels in a site $j$ that is equal to $\mathbf{V}^{(t)}[j,:]\mathbf{1}_{o\times 1}$ to find the location of the voxels to be coloured. This result is further processed to create colour zones, albeit only for illustrative purposes. In other words, we perform all quality evaluations only at the level of the massing configuration rather than the zoning configuration (respectively dubbed \& illustrated as $\boldsymbol{\kappa}$ and $\mathbf{K}$  in our nomenclature). Even in the accessibility score evaluation where we need to know the total volume of each colour in each site, we only project the total colour counts on the single 2D location of the site in question.

\subsection*{Discrete Design Evaluation}
Currently, the game has three evaluation procedures, namely, the total amount of necessary changes of allocation per-site (indicating potentially demolished property volume, possibly of heritage value), annual solar potential,and efficacy of transportation-flows in between the coloured spaces. The latter is chosen to be explained out of these three archetypical evaluation procedures due to its generality and novelty. Without loss of generality, suppose the district provides horizontal access on the ground through some geodesic paths in between the sites and that all coloured spaces are assumed to be projected to a point of entry on the ground as to which their distances are measured (this simplification is necessary for the massing problem at hand and unnecessary for a zoning problem), what is the efficacy of the allocation of colours with respect to the stated preferences for closeness ratings if they are assumed to be estimated transportation flows and the distances considered as transportation costs (as in a transportation problem)? 

\begin{figure}[H]
\centering
\includegraphics[width=16.5 cm]{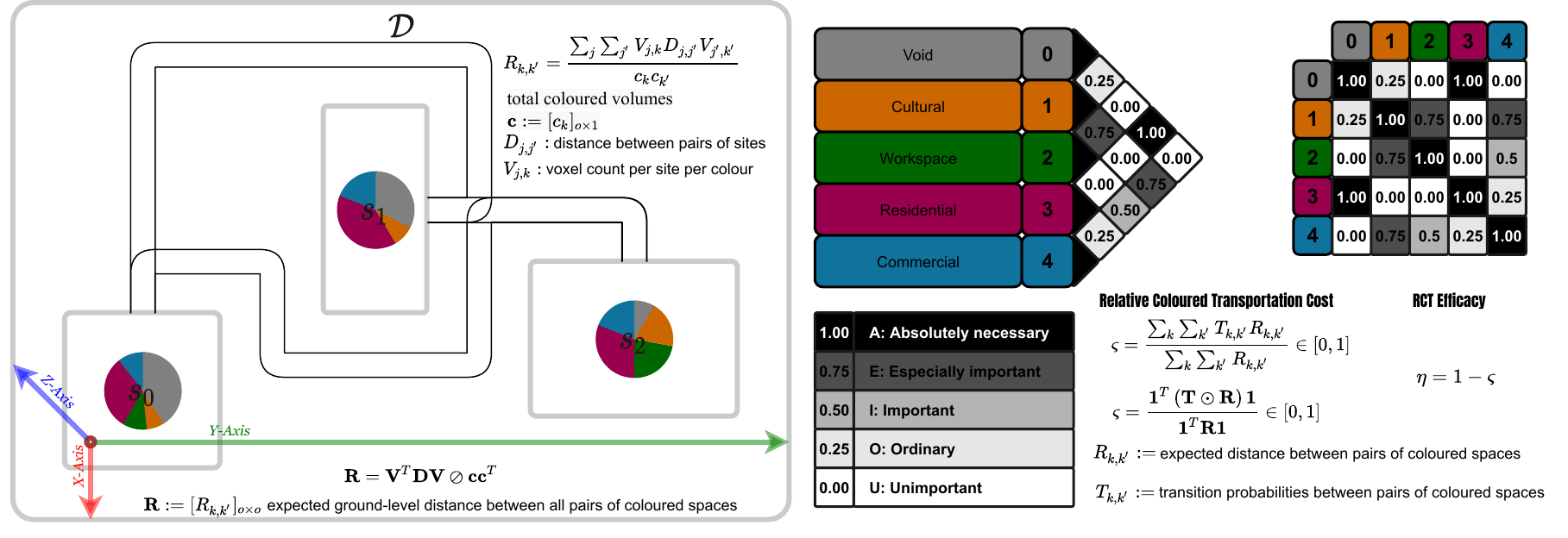}
\caption{Illustrated Coloured Accessibility Evaluation Procedure}
\label{fig:accessibility}
\end{figure}

The answer to this question is formulated in two steps illustrated in Figure\ref{fig:accessibility}: 
Firstly, the relative [expected] distance between the pairs of colours is computed and relativised as to the total amount of coloured spaces present in the allocation-distribution. Secondly, the transportation cost function is formed for the matrix of stated closeness-ratings as a proxy for expected transportation flow-rates between spaces of different colours and the relativised distances between colours are inserted into the equation and the total sum of transportation costs is relativised with respect to the total sum of relativised distances to produce a dimension-less and relative quantity that can be referred to as coloured transportation efficacy, in a manner of speaking. 
Formally, the first part of the procedure is formulated as:
\begin{equation}
R_{k,k'}=
\frac{\sum_{j}
\sum_{j'}V_{j,k}D_{j,j'}V_{j',k'}
}{c_{k}c_{k'}
},
\end{equation}
where $\mathbf{D}:=[D_{j,j'}]_{n \times n}$ denotes the distance between pairs of sites, $\mathbf{V}:=[V_{j,k}]_{n \times o}$ denotes the voxel count per site per colour, and $\mathbf{c} := [c_{k}]_{o \times 1}$ denotes the vector of total colour volumes in the district.
This is summarised algebraically as:
\begin{equation}
\mathbf{R}=\mathbf{V}^T
\mathbf{D}\mathbf{V}\oslash
\mathbf{c}\mathbf{c}^T,
\end{equation}
where $\mathbf{R}:=[R_{k,k'}]_{o \times o}$ denotes the expected ground-level distance between all pairs of coloured spaces. 
Using the relative distances between coloured spaces computed above, we can form a relative transportation cost dubbed $\varsigma\in[0,1]$ to be minimized as to which the efficacy $\eta=1-\varsigma$. Formally, similar to the cost function of a Transportation Problem in Operations Research \cite{lieberman_introduction_2005}, considering the transportation flow-rates $\mathbf{T}:=[T_{k,k}]_{o \times o}$ (stated closeness preferences), we form a cost function relativised by the relative distances:
\begin{equation}
\varsigma=\frac{\sum_{k}\sum_{k'}
T_{k,k'}R_{k,k'}}{\sum_{k}\sum_{k'}
R_{k,k'}}\in [0,1],
\end{equation}
which can be algebraically summarised as:
\begin{equation}
\varsigma=\frac{\mathbf{1}^T
\left (\mathbf{T}\odot\mathbf{R}\right )\mathbf{1}}{
\mathbf{1}^T\mathbf{R}\mathbf{1}}\in [0,1].
\end{equation}

\section*{Implementation \& Experimental Results}

%This section presents the mechanisms underlying the interface of the mathematical game engine with the players for communicating inputs and outputs such as interest and control matrices, weights of quality criteria, preset constraints pertaining to the site, computational procedures proposed for expository qualitative evaluations, and the mathematical definition of the badges used for gamification of the process (Gainer, Player, Contributor).

%An extensive list of libraries and technologies utilized for the implementation of the game is given in the supplementary materials. In short, the back-end of the game has been powered solely by Python (NumPy), and the front-end has been developed in Java-Script React Framework, and the Firebase real-time database on Google Cloud Computing Platform. 

\begin{figure}
    \centering
    \includegraphics[width=16 cm]{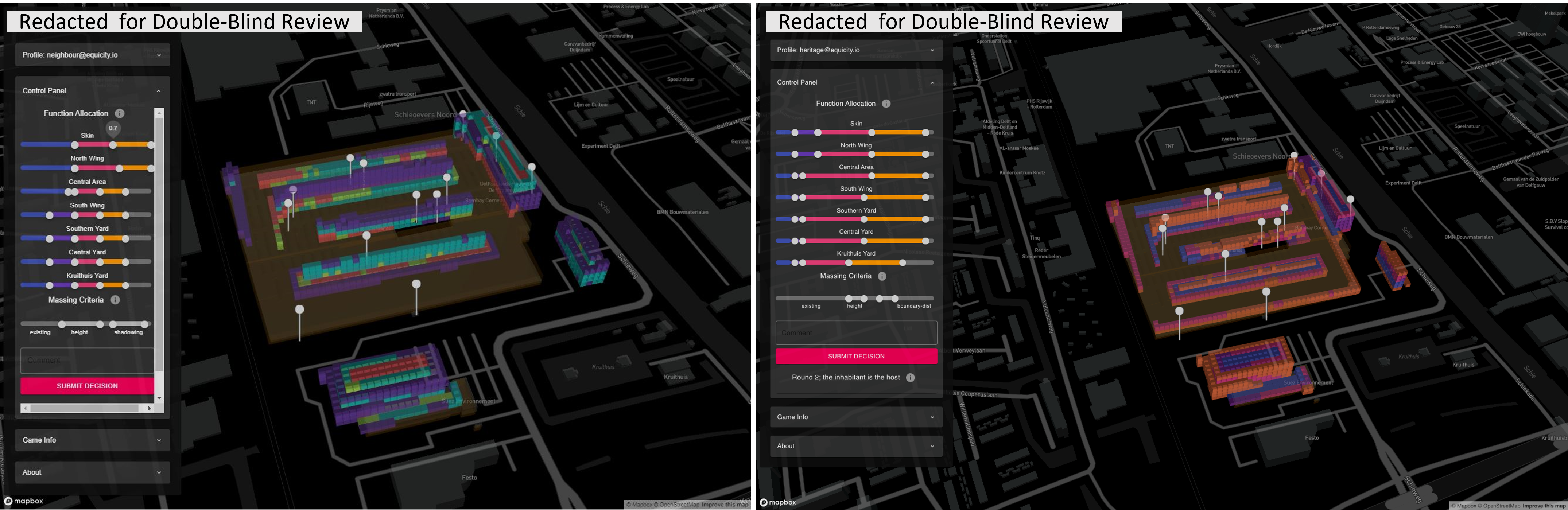}
    \caption{Screenshots of two iterations of game-play, the redevelopment of a former factory into an urban district}
    \label{fig:game_plays}
\end{figure}
The game has been prototyped on a self-developed web-based digital twinning workbench using open source libraries, tools/services, and freely available platforms: \href{https://equicity.emergentium.io/}{https://equicity.emergentium.io} (press Ctrl- upon your visit to adjust the view to your screen).
This section gives an overview of the implemented prototyped game and the typical results obtained through multiple test-play workshops with a group of pseudo-actors (role-players) playing an instance of the game set for contemplating on the hypothetical redevelopment of a former factory into a dense urban neighbourhood. 
The participants were encouraged to focus on the challenge of balancing the conservation of the cultural significance (values \& attributes) of a plot recognized as industrial heritage through the redevelopment process while satisfying the primary objectives of the development, i.e. adding a significant amount of housing units into the district. 
The set up the gaming workshop and the data collected from the workshops are explained in detailed in the supplementary information file. 

The site and the associated fictitious planning problem that were chosen for the development and test cycles are merely to form a vignette for the general idea of participatory generative design of spatial configuration and the challenge of dealing with the idiosyncrasies of the sites particularly in presence of cultural heritage when structuring a systematic participatory design process in the sense of leaving room for negotiations on ad-hoc matters, automating the evaluations that can be automated and providing utmost transparency in the formation and assessment of decisions.  

Further we explain the implementation of the two main components of the game: the interactive interface(front end) and the game engine (back end).
The system architecture described here is that of the prototype tested in the final game-play (test) workshop. 
The workshops required the coordination of a game-master who was to oversee the progress made in terms of negotiation dynamics and attainment of sought qualities. 
Thus, after the final workshop we developed a game master analytic dashboard containing statistical and data-visualization procedures to observe conclusive reports on the decision-making behavior of the players. 

\subsection*{Interactive Interface (Front End)}

\begin{figure}[ht]
    \centering
    \includegraphics[width=13 cm]{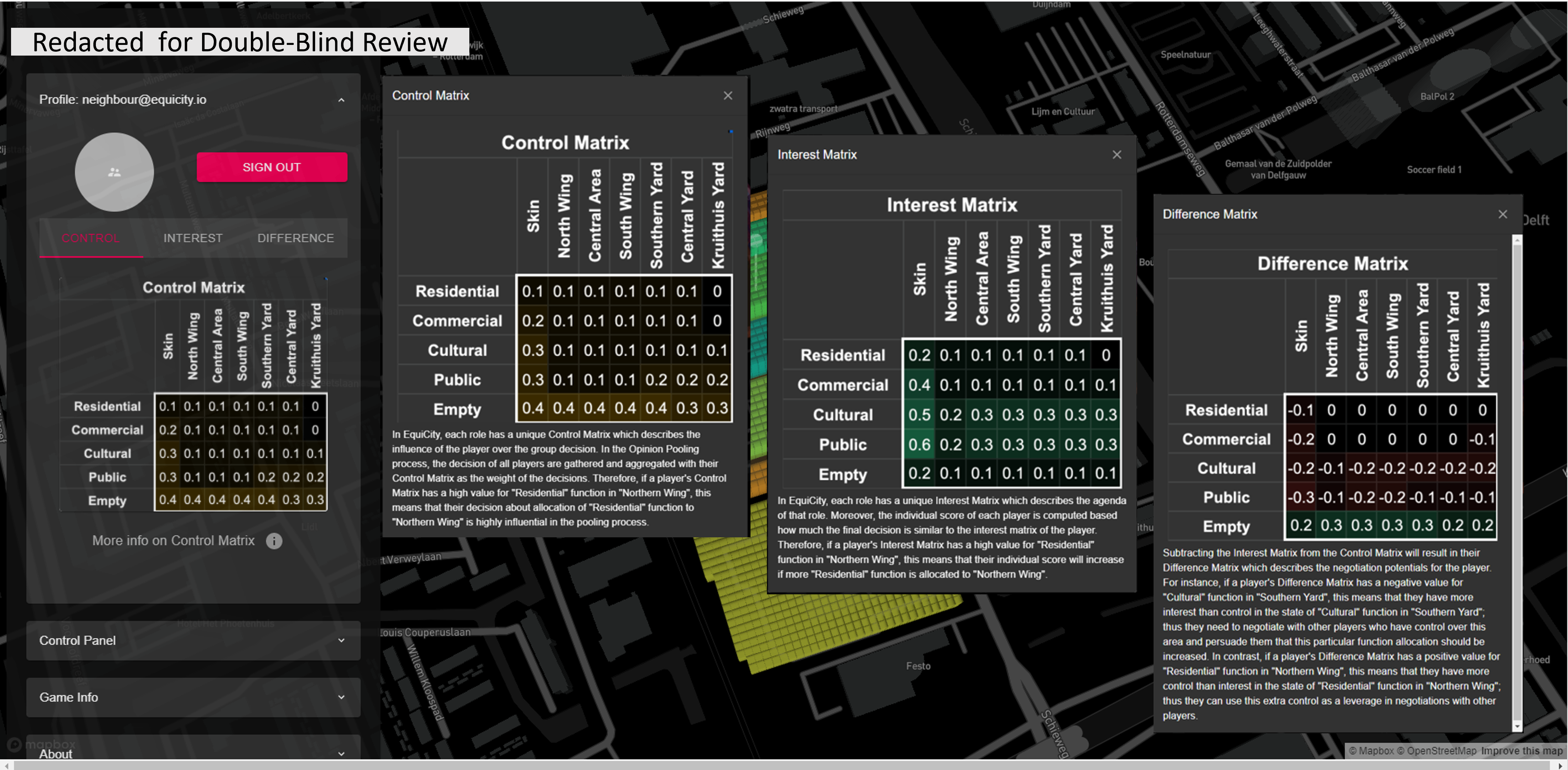}
    \caption{Screenshots of the game interface showing the information panels presented to the players about Control, Interest, and Control-Interest Difference matrices}
    \label{fig:game_settings}
\end{figure}

\begin{figure}
    \centering
    \includegraphics[width=13 cm]{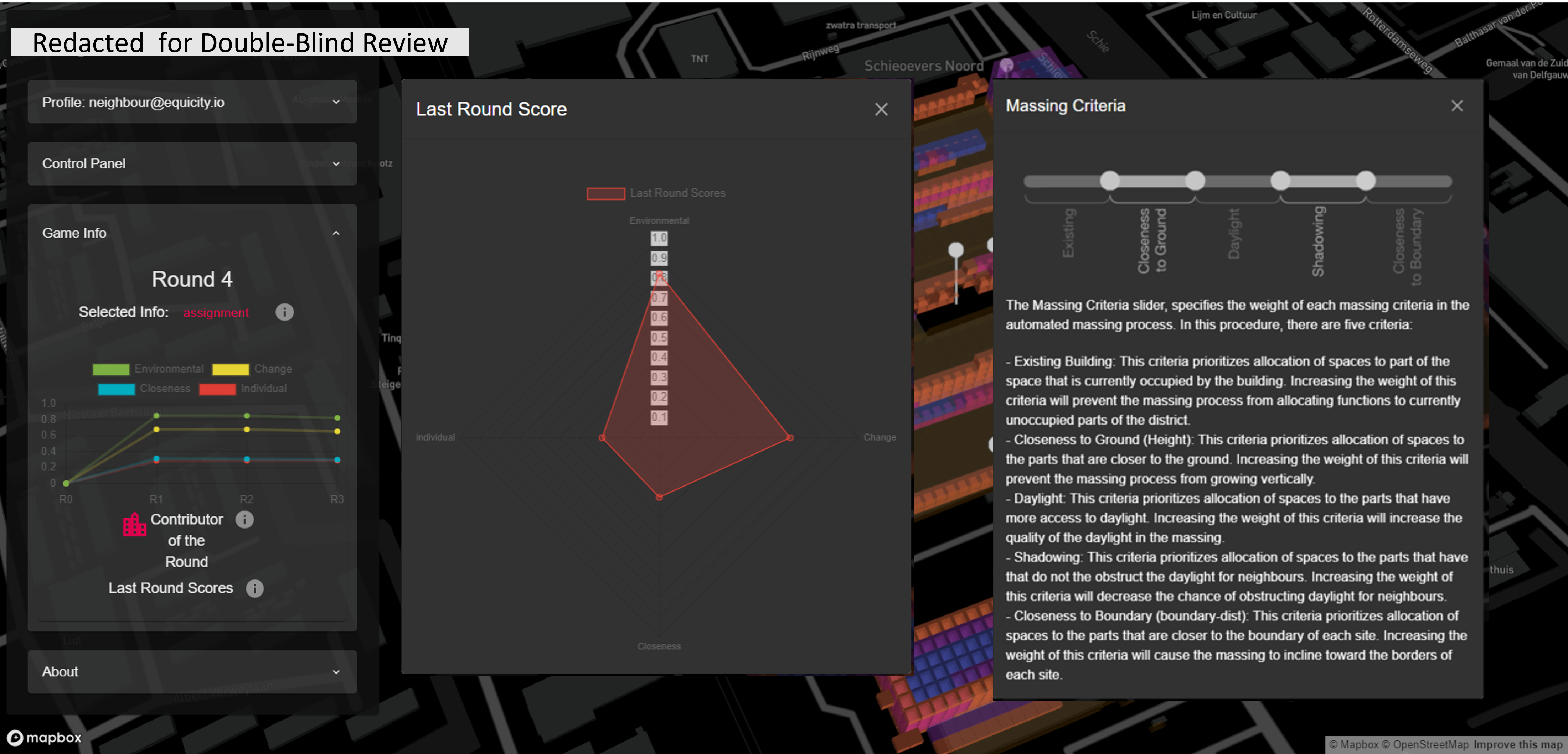}
    \caption{Screenshots of the evaluation output interface of the game and the information provided to players on how to adjust the weights of massing criteria}
    \label{fig:game_scores}
\end{figure}

The interactive interface allows the players to explore spatial information, environmental analyses, their individual scores \& badges, the agenda of their roles (i.e. the interest matrix at time 0, dubbed $\mathbf{X}^{(0)}$ that is supposedly the initial agenda or the mindset of the entity on behalf of whom they act as agents or proxies), and most importantly, to express their decisions for the next round (interests $\mathbf{X}^{(t)}$ and weights $\mathbf{W}^{(t)}$ for the massing criteria).
The interface is web-based and so it does not require any prior installation for the participants.
Everyone can visit the website to observe the game-play session while it is running and interactively explore the additional information such as spatial analysis and group scores. 
Each player needs to login with their credentials to access their agenda specified as `control, interest, and difference matrix (later referred to as the surplus matrix in the text for notational consistency)' (cf. Figure \ref{fig:game_settings}); make decisions in terms of allocation of colors to sites; and check their individual scores and badges as visible in Figure \ref{fig:game_scores}.

The front end is implemented using React framework in Java-Script;
the maps were added using MapBox;
geospatial information were visualized using Vis.gl \cite{noauthor_visgl_nodate};
and finally scores were visualized using D3 \cite{bostock_d3_2011}.

Within the game-play, in each round, the players input their decisions regarding the allocation of colors to each site through the available sliders in the control panel and specify the weights of massing criteria and add a comment describing the motives behind their decision.(visible in Figures \ref{fig:game_settings}\ref{fig:game_settings}\ref{fig:game_scores}).
Once satisfied with the negotiations, the players submit their decision and await other players to submit their decisions.
Once all decisions are submitted, the game engine kick-starts and goes through a cycle of opinion pooling, proportional fitting, massing, and evaluation.
When the cycle is finished all the proceedings are updated in the database and the interface shows the results.
These updated data include the spatial distribution of the voxel values, the new mass configuration, the previous massing, submitted decisions (interests, weights), updated scores and badges.

Players, at any time during or after the session, can access aggregated information through the profile and game info menu. 
Additionally, they can visualize various contextual information and explore the district in the integrated 3D environment, e.g. they can access the annotated heritage attributes and values of the site with their corresponding images and texts by hovering over the pins scattered over the site.
%Moreover, the interface provides extra information and description to the players in the format of a small \emph{i} button that opens a pop-up window containing the supplementary information. 

%\FloatBarrier
\subsection*{Game Engine (Back End)}

The Game Engine includes all of data processing functionalities of the system.
Once all decisions are submitted the engine is triggered to perform the following: 

First the individual decision of the players ($\mathbf{X}^{(t)}_{m \times n \times o}$) are gathered from the database; 
the opinion pooling is performed to produce the collective decision $\mathbf{A}^{(t)}_{n \times o}$; 
the iterative proportional fitting is performed to produce the Volume-Decision matrix $\mathbf{V}^{(t)}_{n \times o}$; 
the massing is performed to achieve the volumetric configuration of each site $\mathbf{\kappa}$;
then the evaluation of spatial massing quality criteria is performed to produce $\mathbf{q}_{e \times 1}$;
the procedure for allocating game badges is executed;
and finally all of the information including the decision variables, evaluation fields $\varphi$, aggregated scores and game-play badges.
The disaggregated environmental evaluations are computed per voxel $\varphi_l$ a priori. The evaluation at the aggregate level to produce the vector of multi-criteria outcomes of interest ($\mathbf{q}$) is thus only a matter of spatial aggregation (integration).
Explaining the exact procedures for computing $\varphi_l$ would go far beyond the scope of this paper and so only one important example of aggregate quality criteria is presented in the paper (coloured accessibility evaluation). 

One of the complicated aspects of the game is that the lists of aggregate outcomes of interest and disaggregated fields of quality do not have a one to one correspondence, e.g. the aggregate accessibility score is not disaggregated to lead the massing process. 
In other words, the criteria used for massing are not exactly the same as those displayed as scores. 
Addressing this issue would fall far outside of the scope of this paper. 
However, the framework as presented here is to structure such processes and so, without loss of generality, we claim that the proposed examples are generic enough to be representative of the bigger idea of structuring such a complex group decision-making process and nudging it towards sustainable and equitable outcomes. 

The game engine is implemented in python utilizing the following open source libraries:
\emph{NumPy} \cite{harris_array_2020} for implementing the algebraic processes;
\emph{Pandas} \cite{noauthor_pandas-devpandas_nodate,mckinney_data_2010} for structuring and organizing data;
\emph{topoGenesis} \cite{azadi_topogenesis_2021} for spatial indexing and topological functionalities of the volumetric units;
and finally \emph{HoneyBee} \cite{roudsari_automate_2016,roudsari_ladybug-toolshoneybee_2021} and \emph{EN 17037 Recipes} \cite{brembilla_computational_2021} for performing visibility and solar analyses.

% \begin{table}[h]
%     \caption{Meta-procedures in the proposed framework}
%     \footnotesize
%     \begin{center}
%         \begin{tabularx}{\hsize}{
%                 >{\footnotesize\centering\arraybackslash\hsize=0.9\hsize}X
%                 >{\footnotesize\centering\arraybackslash\hsize=0.9\hsize}X
%                 >{\footnotesize\centering\arraybackslash\hsize=0.9\hsize}X
%                 >{\footnotesize\centering\arraybackslash\hsize=1.3\hsize}X
%             }
%             \toprule
%             \textbf{Process}
%             & \textbf{Product}
%             & \textbf{Scope}
%             & \textbf{Complexities}
%             \\
%             \toprule				
%             Planning
%             & Network
%             & Graph Theory
%             & multi-value \linebreak multi-actor \linebreak multi-criteria
%             \\
%             \midrule
%             Configuring
%             & Configuration
%             & Algebraic Topology
%             & multi-dimensional \linebreak multi-criteria 
%             \\
%             \midrule
%             Shaping
%             & Shape
%             & Geometry
%             & multi-dimensional \linebreak multi-value 
%             \\
%             \toprule
%         \end{tabularx}
%     \end{center}
    
%     \label{tab:framework}
% \end{table}

\section*{Discussion}
The paper puts multi-actor consensus and multi-criteria optimization in a challenging spatial context of a design \& planning problem. Such problems are notoriously difficult to formulate and tackle, hence the term 'wicked problems' \cite{rittel_dilemmas_1973} commonly used to refer to them. In this endeavour, we went beyond such vague notions and addressed the complexity of a generic class of such problems by structuring it mathematically in a novel, straightforward, and open-ended framework without compromising its multiplex sophistication. 
The algebraic structure of the proposed framework not only makes it elegant and easily explainable, but also very efficient for large-scale implementations that could massively scale-up participatory decision-making processes.
Without claiming that all the multiple facets of multi-criteria decision making have been addressed adequately in our experiments, we invite the scientific community to utilize the framework and test the efficacy of various forms of structured group decision-making, especially with a focus on balancing the importance of inter-subjective consensus and optimality w.r.t. objective quality criteria. 

With hindsight, following our reflections on the proceedings of the game-play workshops, we were able to identify the contributions and limitations of the proposed serious gaming framework and avenues for further research for generalization of the game to more complex settings or adaptation of the game to other types of spatial or non-spatial Multi-Actor Multi-Criteria Decision Making problems. Namely, in spite of the aim of the project to explicitly address the notion of equity, we did not quantify it, and yet we can envisage that the  
%The Discussion should be succinct and must not contain subheadings.
game-badges can be developed further to produce the equity scores. Considering that equity is eventually about fairly sharing both the costs and the benefits of developments, budgeting and cost-sharing definitely need to be integrated in the future developments.
The implicit meaning of not having considered costs for proposed developments is that the players are acting as idealistic ``do-gooders'' without worrying about the costs of what they are proposing to be developed. While addressing this issue may be rather straightforward mathematically, incorporating it in the narrative of the game meaningfully would require much effort to ensure consistency. 
Another generalization that is needed is the consideration of insistence of some agents on the initial agendas, similar to the formulation of Friedkin et al \cite{friedkin_mathematical_2019}.
The game is arguably successful in featuring an accessible and scaleable participatory design mechanism. The proposed social choice mechanism allows for simulating otherwise long processes of consensus building by automatically going through rounds of iterations to converge to a point of equilibrium. This effectively allows the players to use the time of the session more effectively for building different kinds of consensual decisions and reflect on the consequences of their choices in the bigger scheme of the neighbourhood and sustainable development goals.

The results of the particular test-case scenario (which is fictitious) suggest that there are some loose ends in particular with regards to the explainability of evaluations and the generation of zones; however, these processes are merely illustrative of the more general idea of a configurator as a gamified "play \& score" mechanism. 
%\bibliography{EquiCityLib}
One point of improvement concerns the consideration of MCDA results in issuing the gamification badges of honour. 
A player might have been very cooperative or competitive in reaching a consensus that is respectively close to their positive power surplus or negative power surplus but that consensual decision may not have necessarily resulted in good outcomes from a multi-criteria decision analysis point of view. In fact, as expected, one of the most difficult (abstract) aspects of the game for the players is to guess the complex associations between their decisions and their measured outcomes (objective functions). 
This complexity cannot necessarily be alleviated by explaining the advanced settings of the system or the mathematical process, since it pertains to the domain-specific physics of each matter being addressed by each one of the performance/quality indicators. 
No single player can be expected to be familiar with all of the concepts behind these specific evaluation modules. In fact, the domain-specific knowledge embedded in each one of these evaluation modules is what can be referred to as engineering design expertise, i.e. the know-how rooted in the associations between the decision-variables related to shape and configuration on the one hand and the quality, performance, or functionality of the outcome on the other hand. 

The associations between the configurations and their objectively measured qualities (the so-called Key Performance Indicators) are often notoriously baffling to comprehend for even novice or intermediate designers, let alone lay players of the game. 
Notwithstanding the cognitive difficulty of making the right decisions, the point of devising and playing such a game is to explicate the problem as a decision-making problem to provide direct control for whomever has a stake in the state of the object being designed, be it a neighbour or a prospective inhabitant of the area. 
However, further contemplation is needed on the distinction between stakeholders who will have to abide by their own decisions and financially partake in the implementation of the decisions and the actors who might have a stake in the development of the site as neighbours (out of the theoretical system). 
The point is, it is easy for people who do not have to bear the costs of supposedly good decisions to vote for the most progressive options but nudging the financial stakeholders towards sustainable choices in a persuasive manner is quite a different challenge altogether.

\section*{Data Availability}
The datasets generated and/or analysed during the current study are available in the EquiCityData repository \\ \href{https://github.com/shervinazadi/EquiCity_Data}{https://github.com/shervinazadi/EquiCity\_Data}.

%The data that has been generated or analysed during the current study are available on a \href{https://github.com/shervinazadi/EquiCity_Data}{GitHub repository}. 
%Algorithms, analyses, and important derivations have been reported separately in the Supplementary Information Document. 
%Additional information about the project can be found on \href{https://genesis-lab.dev/products/equicity/}{this page}. 

\bibliography{EquiCity_Nature}

%\noindent LaTeX formats citations and references automatically using the bibliography records in your .bib file, which you can edit via the project menu. Use the cite command for an inline citation, e.g. 

For data citations of datasets uploaded to e.g. \emph{figshare}, please use the \verb|howpublished| option in the bib entry to specify the platform and the link, as in the \verb|Hao:gidmaps:2014| example in the sample bibliography file.

\section*{Acknowledgements}
The \href{https://www.nwo.nl/en/projects/nwa1228192313}{EquiCity} project has received funding from the Dutch National Science Foundation (NWO), the funding shceme \href{https://www.nwo.nl/en/researchprogrammes/dutch-research-agenda-nwa/innovation-and-networks-nwa/idea-generator-nwa-idg-0}{Idea Generator} (NWA-IDG) under grant agreement No. NWA1228192313. Additionally, we would like to thank Ms. Ilaria Rosetti, Dr. Marzia Loddo, Mr. Jos Wassing, Ir. Kotryna Valeckaite, Ir. Aditya Pravin Soman, Dr. Michael van der Meer, Ms. Maartje Damen, Ms. Federica Romagnoli, Ms. Deniz Erdem Okumuş, Ms. Vasilka Espinosa, Mr. Sebastian Fischer Stripp, Ms. Eda Akaltun, Ir. Simon Tiemersma, Ir. Michiel Susebeek, Ms. Mahda Foroughi, Ms. Ana Tarrafa da Silva, and Ir. Pien Tol who participataed actively as role-players in our three test-play workshops during the project for their remarkable committment and providing constructive feedback on the early versions of the game.

\section*{Competing Interests}
The author(s) declare no competing interests. 
%Acknowledgements should be brief, and should not include thanks to anonymous referees and editors, or effusive comments. Grant or contribution numbers may be acknowledged.

\section*{Author contributions statement}

%PZN \& SAZ conceived the work, devised the mathematical process, and wrote the first prototypes of the game engine. PZN, SAZ \& APR acquired the funding.  PZN \$ SRZ wrote the first manuscript. SAZ \& NAZ developed the Front-End interface of the game. SAZ \& PZN developed the Back-End of the game; SAZ integrated, developed, and generalized the platform. NBI and SAZ conducted the statistical validation tests for devising the Game Master's dashboard and analysed the results. BDA, SAZ, \& PZN gamified the mathematical design process together. 
Contributions stated according to the \href{https://casrai.org/credit/g}{CRediT – Contributor Roles Taxonomy}:

\textbf{PZN}: Conceptualization, Methodology, Formal analysis, Software, Visualization, Writing- original draft preparation, Funding acquisition, Investigation, Supervision, Project administration, Validation. 
\textbf{SAZ}: Conceptualization, Methodology, Formal Analysis, Software, Data curation, Visualization, Funding Acquisition, Resources, Writing – review \& editing. 
\textbf{NBI}: Validation, Formal analysis, Investigation, Writing – review \& editing. 
\textbf{BDA}: Conceptualization, Data curation, Investigation.
\textbf{NAZ}: Software, Resources. 
\textbf{SRZ}: Conceptualization, Writing- original draft preparation
\textbf{APR}: Conceptualization, Supervision, Funding acquisition.
All authors reviewed the manuscript.
%\textcolor{red}{All authors reviewed the manuscript.}
%Must include all authors, identified by initials, for example:
%A.A. conceived the experiment(s),  A.A. and B.A. conducted the experiment(s), C.A. and D.A. analysed the results.   

\section*{Additional information}

To include, in this order: \textbf{Accession codes} (where applicable); \textbf{Competing interests} (mandatory statement). 

The corresponding author is responsible for submitting a \href{http://www.nature.com/srep/policies/index.html#competing}{competing interests statement} on behalf of all authors of the paper. This statement must be included in the submitted article file .

% \begin{figure}[ht]
% \centering
% \includegraphics[width=12 cm]{stream}
% \caption{Legend (350 words max). Example legend text.}
% \label{fig:stream}
% \end{figure}

% \begin{table}[ht]
% \centering
% \begin{tabular}{|l|l|l|}
% \hline
% Condition & n & p \\
% \hline
% A & 5 & 0.1 \\
% \hline
% B & 10 & 0.01 \\
% \hline
% \end{tabular}
% \caption{\label{tab:example}Legend (350 words max). Example legend text.}
% \end{table}

%Figures and tables can be referenced in LaTeX using the ref command, e.g. Figure \ref{fig:stream} and Table \ref{tab:example}.

\end{document}

\maketitle
%\flushbottom
\renewcommand{\figurename}{Supplementary Figure}
\renewcommand{\tablename}{Supplementary Table}

\thispagestyle{empty}

\section*{Glossary}
\subsection*{Nomenclature}
Here we present a summary of the essential notations used in the remainder of paper in table \ref{tab:nomenclature}. Our notational conventions are mostly revolving around the paradigm of vector-computing (algebraic code using vectors, matrices, and tensors), as to which most element-wise expansions and summation notations are omitted for the sake of brevity. We use lower case Latin and Greek letters for scalars, bold lower-case Latin and Greek letters for vectors, bold Latin, Greek, Fraktur, and Cyrillic letters for matrices and tensors, use the slicing 'colon notation' of Golub and van Loan \cite{golub_matrix_2013} (consistent with NumPy syntax) for denoting matrix/tensor rows, columns, or pages (e.g. $\mathbf{X}[:,:,k]$ denotes the $k_{th}$ page of the tensor $\mathbf{X}$). Dot products are omitted and Hadammard-Schur (element-wise) products/divisions are denoted respectively with $\odot$ and $\oslash$. Augmenting (concatenating/appending) matrices by vectors/matrices is denoted by $|$, e.g. $[\mathbf{A},\mathbf{b}]$ implies that the vector $\mathbf{b}$ is appended to the right side of the matrix $\mathbf{A}$. Unless otherwise stated, all vectors are assumed to be column-vectors. 

\begin{table}
\centering
\resizebox{\columnwidth}{!}{%

\begin{tabular}{|l|l|l|}
\hline
Symbol & Data Type/Shape & Description \\
\hline
$i$ & integer index of actors & actors: ${i\in[0,m)}\subset\mathbb{N}$ \\
\hline
$j$, $j'$ & integer index of sites & sites: ${j\in[0,n)}\subset\mathbb{N}$ \\
\hline
$k$, $k'$ & integer index of colors & colours: ${k\in[0,o)}\subset\mathbb{N}$ \\
\hline
$\mathbf{X}$ & $\mathbf{X}:=[X_{i,j,k}]_{m\times n \times o}$, a matrix of floats in range of $[0,1]$ & Interest Matrix (actors, sites, and colors), no time-stamps for brevity \\
\hline
$\mathbf{C}$ & $\mathbf{C}:=[C_{j,i,k}]_{n\times m \times o}$, a matrix of floats in range of $[0,1]$ & Control Matrix (sites, actors, and colors), fixed in advanced settings \\
\hline
$\mathbf{A}^{(t)}$ & $\mathbf{A}^{(t)}:=[A_{j,k}]_{n\times o}$, matrix of floats in $[0,1]$ at time $t$& Collective Decision Matrix, desired areas for sites (sites and colors) \\
\hline
$\mathbf{V}^{(t)}$ & $\mathbf{V}^{(t)}:=[V_{j,k}]_{n\times o}$, matrix of positive integers at time $t$ & Volume-Decision Matrix, desired volumes for sites (sites and colors) \\
\hline
$\mathbf{c}$ & $\mathbf{c} := [c_{k}]_{o \times 1}$, vector of positive floats & District Level Program as Required Surface-Area per Color \\
\hline
$\mathbf{T}$ & $\mathbf{T} := [T_{k,k'}]_{o \times o}$, matrix of positive floats in $[0,1]$ & Activity Relations Chart: closeness ratings between coloured spaces \\
\hline
\end{tabular}%
}
\caption{\label{tab:nomenclature}A summary of the essential notations used in the paper}
\end{table}
\begin{itemize}
    \item \textbf{Actors}: define the local tendencies for coloring sites
    \item \textbf{Sites}: need to accommodate all the allocated colors
    \item \textbf{Colours}: represent the discrete types of uses, each with a given surface area from the district-level programme
\end{itemize}
\subsection*{Definitions}
Here we give a brief explanation about the concepts used in this report: 
\subsubsection*{Area Development vs Gentrification}
    Sometimes a euphemism for a gentrification process, area development usually refers to the development of less-urban areas (such as former military or industry zones) typically surrounded by newer urban developments into new neighbourhoods, mostly incorporating many sought-after housing units and, if regulations, policies, or activism succeeds, a mix of communal spaces, public spaces, cultural spaces, and commercial spaces. From the point of view of the game an urban area development problem, without loss of generality, is regarded as a spatial investment portfolio-management problem.
\subsubsection*{Actor vs Stakeholder}
An actor is a real/legal person or their proxy that is assumed to be playing the participatory design game. Not everyone who has a stake in the development can be guaranteed to be an actor nor vice-versa. A stake-holder would also get a supposedly fair share of both the costs and the benefits of a certain development. However important this may be for the future generalizations of the game, we have not taken the cost-sharing of the proposed developments into account in the scope of the presented game. Thus the players are referred to as actors, who may or may not have the concern of bearing the costs of developments they propose.
\subsubsection*{Interest vs Control} 
An actor may have various levels of financial/personal interest in a certain type of investment (e.g. housing, cultural, commercial, et cetera) but not necessarily the same level of control in terms of power over the final decision over the amount of investment on their matter of interest. 
\subsubsection*{Opinion Pooling} 
A systematic process of consensus-building (a generalized extension of the process introduced by Batty \cite{batty_evolving_2016}) in which a point of equilibrium is made iteratively by averaging the votes of the actors for investments on a diverse portfolio of distinct investment objects (figuratively referred to as colours/labels in the game) in a number of sites (could be multiple investment portfolios in a group decision-making on resource allocation as referred to in \cite{friedkin_mathematical_2019}). This procedure is a Markovian Design Machine as introduced by Batty \cite{batty_theory_1974} that functions as a consensus-building procedure (as a basis for a consensus on the voting procedures, referred to as procedural consensus in Friedkin's book on Social Influence \cite{friedkin_structural_1998}).
\subsubsection*{Iterative Proportional Fitting}
In various use-cases such as statistical estimation of population distributions or, as identified in this paper, in planning procedures where there is an aggregate picture of a distribution and one needs to work out a disaggregated tabular picture of a distribution to agree in row-sums and column-sums with the prescribed or expected row-sums and column-sums a procedure dating back to 1940's works of statisticians Stephen \cite{stephan_iterative_1942} \& Deming \cite{deming_least_1940}, nowadays known as Iterative Proportional Fitting \cite{hunsinger_iterative_2008} is used to adjust the matrices to ensure that the local disaggregated distribution conforms to the expected/prescribed global aggregate picture. 
\subsubsection*{Generative Massing vs Zoning}
Generative Design is an umbrella term that refers to simulation-driven processes of computational design formulated as discrete decision-making/optimization problems (mostly shape or topology optimization) that concern the topology/configuration of a spatial object in terms of a binary mass-void distribution. When a single such distribution/configuration is concerned we can refer to it as a massing problem (\emph{vide infra}). If multiple such distributions are to be concerned simultaneously, one has to deal with a 'coloured' distribution (in a graph-theoretical sense colours are in fact distinct categorical labels represented as integers). The latter type of colourful distribution is referred to as zoning in our jargon. Whilst zoning [categorical] distributions are shown in the paper for illustrative purposes, the qualitative evaluation of a zoning distribution falls out of the scope of the paper. 
\subsubsection*{Performance Evaluation vs Quality Assessment} 
While the two terms are practically synonymous, the connotation of the former in the literature is related to the so-called high-performance engineering and objective measurement of physical quantities, while the latter is mostly associated with problems involving human factors and uncertain decision-making, thus it is preferred in this context. It must be noted that the kind of assessment performed in the game must be \emph{ex-ante assessment} of potential outcomes based on a priori know, and not \emph{ex-post assessment} of actual outcomes, i.e. a posteriori. Given a representation of the designed urban neighbourhood as a mass-void distribution, typically referred to as an envelope, we can estimate the performance of a hypothetical building complex bound into this envelope e.g. with respect to the solar potential of the building or the efficacy of the spatial distribution of activities with reference to an Activity-Relations-Chart representing preferred closeness ratings between the constituent spaces of the neighbourhood. The presented evaluation criteria are merely there to exemplify a broad variety of other such criteria that could be considered for evaluation mass (black \& white) or zone (coloured) configurations. 
\subsubsection*{Multi-Objective Optimization vs Multi-Criteria Decision Analysis}
\begin{figure}
    \centering
    \includegraphics[width=13.5 cm]{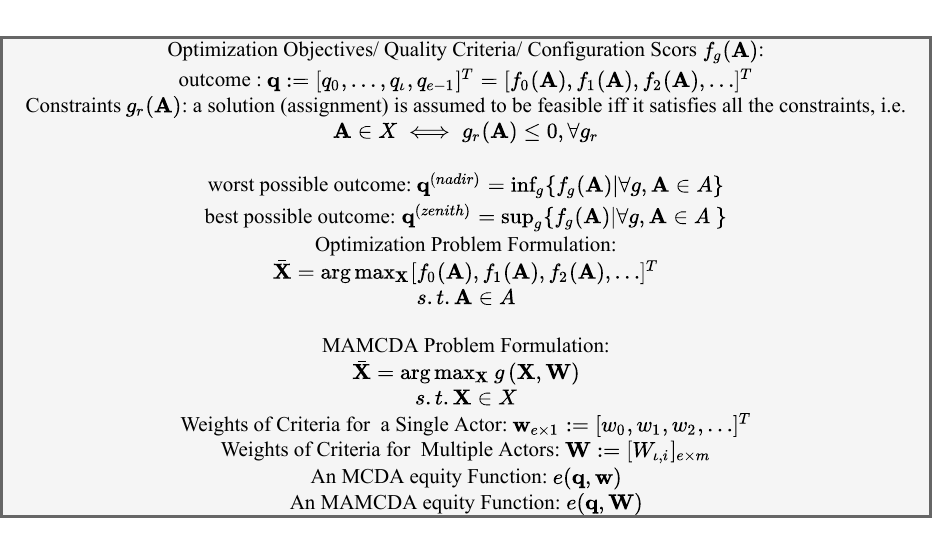}
    \caption{Illustrated formulation of the problem as a gamified negotiation problem}
    \label{fig:MOO-vs-MCDM}
\end{figure}

Since the distinction between the connotations two terms is as wide as the gap between their application areas in physics-dominated engineering and Operations-Research/Management-Science, one cannot easily find a clear-cut distinction between the two in the literature. However, for avoiding any confusion, we find it important to state the subtle difference as addressed by H.A. Simon \cite{simon_administrative_1997}, Jackson \cite{jackson_mechanism_2014}, and Hwang et al. \cite{hwang_group_1987}, is that single objective optimization (also known as programming, as in Linear Programming, Integer Programming, and Quadratic Programming in Operations Research) or multi-objective optimization formulation of decision making problems is only possible/meaningful in presence of objectivity and existence of a uniform view on the nature of the objectives and measurements/estimations of attributes as to which the objectives are assessed. When there are multiple actors involved, the uncertainties arising from lack of data, the differences of the value-systems of actors, and the game-dynamics of group decision making complicate the process of finding satisfactory decision-alternatives. Hwang and Lin have explained this briefly, by tangentially referring to the rejection of non-dominated solutions in the sense of Pareto-Optimality criterion:
    "Moving from a single decision maker to a multiple decision maker setting introduces a great deal of complexity into the analysis. The problem is no longer the selection of the most preferred alternative among the non-dominated solutions according to one individual's (single decision maker's) preference structure. The analysis must be extended to account for the conflicts among different interest groups who have different objectives, goals. criteria. and so on \cite{hwang_group_1987}."
See a comparative formulation of the posed problem in MOO and MCDM form in Figure \ref{fig:MOO-vs-MCDM}.

\begin{figure}
    \centering
    \includegraphics[width=16.5 cm]{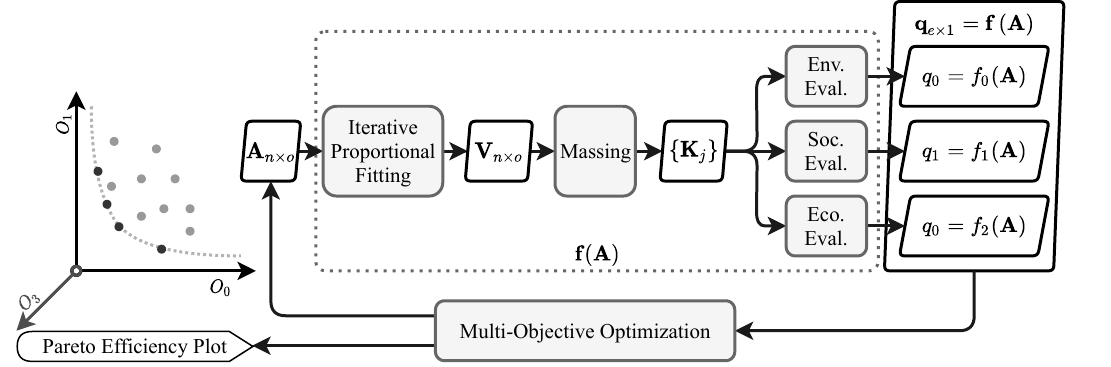}
    \caption{Illustrated formulation of the problem as an optimisation problem (not solved in this paper, \emph{vide infra} the gamified negotiation problem)}
    \label{fig:optimization-problem}
\end{figure}

\FloatBarrier

\section*{Method}

The following figure \ref*{fig:dataflow} gives an verview of the proposed methods and the way they constitute the proposed generative design game:

\begin{figure}[H]
    \centering
    \includegraphics[width=15 cm]{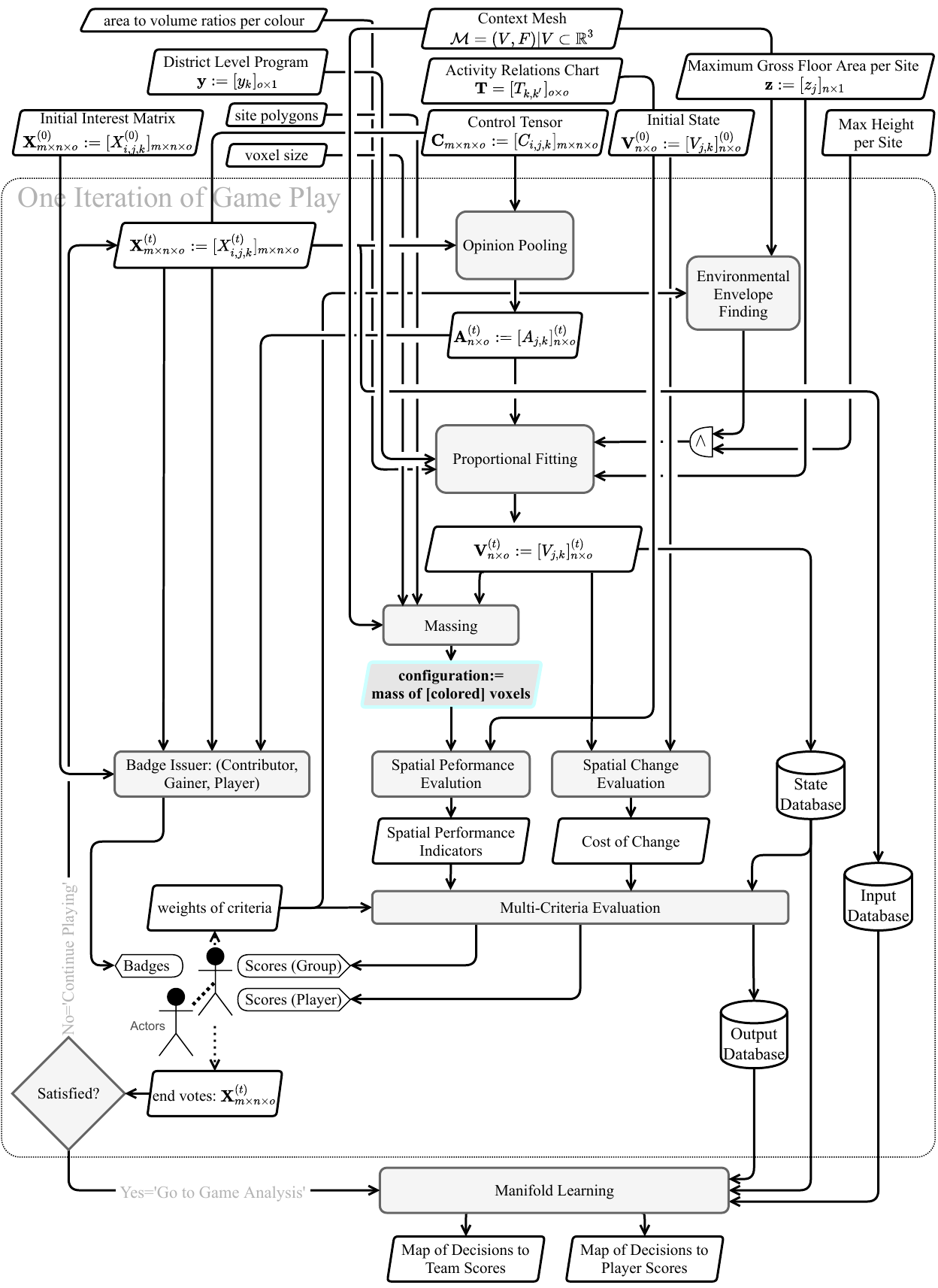}
    \caption{Data-Flow of the Proposed Participatory Design Game}
    \label{fig:dataflow}
\end{figure}

\subsection*{Derivation of Gamification Badges}

The Winner is the actor with the most similar  interest matrix to
the aggregated decision, formally:
\begin{equation}
    \arg\min_{i} \|\mathbf{X}^{(t)}[i,:,:]-\mathbf{A}\|_{F}.
\end{equation}
%$$\arg\min_{i} \|\mathbf{X}^{(t)}[i,:,:]-\mathbf{A}\|_{F}.$$

The Loser is the actor with the most dissimilar  interest matrix to
the aggregated decision, formally:
\begin{equation}
    \arg\max_{i} \|\mathbf{X}^{(t)}[i,:,:]-\mathbf{A}\|_{F}.
\end{equation}

Out of these two badges, only the first one is communicated by the game engine as the [strategic] gainer of the round.
In addition, we propose two more sophisticated badges based on a definition of negotiation power (not dissimilar to the one introduced by Steunenberg et al \cite{steunenberg_strategic_1999}):

We define a Power Surplus Matrix for the $i_{th}$ actor as:
\begin{equation}
    \boldsymbol{\Pi}^{(t)}_i:=\mathbf{C}[:,i,:]-\mathbf{X}^{(t)}[i,:,:].
\end{equation} 

The Frobenius Norm of the tensor made up of the stack of  all such matrices $\|\boldsymbol{\Pi}^{(t)}\|_F$ indicates the total motivation for negotiations during the game between those actors with much interest and little control on some investment and those with much control and little interest:
\begin{equation}
    \boldsymbol{\Pi}^{(t)}:=\mathbf{C}^{T(1,0,2)}-\mathbf{X}^{(t)}
\end{equation}

The  badge of honour "Player of the Round" is defined as
the actor with the most similar pattern of the negative parts of their Power Surplus matrix to
the aggregated decision, or formally as:
\begin{equation}
\arg\min_{i} \left[\pi_{\ominus}^{(t)}[i]\right]_{m\times1}
\end{equation}
, where
\begin{equation}
\boldsymbol{\pi}_{\ominus}^{(t)}:=[\pi_{\ominus}^{(t)}[i]]_{m\times 1}=
\left(
\left(\left(\boldsymbol{\Pi}^{(t)}_{\ominus}\odot \mathbf{A}^{(t)}\right)_{m\times n \times o}\mathbf{1}_{o \times 1}\right)\mathbf{1}_{n\times 1}\right)\oslash
\left(\left(\boldsymbol{\Pi}_{\ominus} \mathbf{1}_{o \times 1}\right)\mathbf{1}_{n\times 1}\right)
\end{equation}
in which
\begin{equation}
\boldsymbol{\Pi}_{\ominus}:=
-\left(\boldsymbol{\Pi}<0\right)\odot\boldsymbol{\Pi}.
\end{equation}
The badge of honour "Contributor of the Round" is defined as
the actor with the most similar pattern of the positive parts of their Power Surplus matrix to
the aggregated decision, or formally as:
\begin{equation}
\arg\min_{i} \left[\pi_{\oplus}^{(t)}[i]\right]_{m\times1}
\end{equation}
, where
\begin{equation}
\boldsymbol{\pi}_{\oplus}^{(t)}:=[\pi_{\oplus}^{(t)}[i]]_{m\times 1}=
\left(
\left(\left(\boldsymbol{\Pi}^{(t)}_{\oplus}\odot \mathbf{A}^{(t)}\right)_{m\times n \times o}\mathbf{1}_{o \times 1}\right)\mathbf{1}_{n\times 1}\right)\oslash
\left(\left(\boldsymbol{\Pi}_{\oplus} \mathbf{1}_{o \times 1}\right)\mathbf{1}_{n\times 1}\right)
\end{equation}
in which
\begin{equation}
\boldsymbol{\Pi}_{\oplus}:=
+\left(\boldsymbol{\Pi}>0\right)\odot\boldsymbol{\Pi}.
\end{equation}

Note that the input of the badge issuer is the matrix $\mathbf{A}^{(t)}$ and not the matrix $\mathbf{V}^{(t)}$. 
This is because the matrix $\mathbf{A}^{(t)}$ is comparable with $\mathbf{X}^{(t)}$, $\mathbf{X}^{(0)}$, and $\mathbf{C}$ in that all of these matrices contain percentages, but $\mathbf{V}^{(t)}$ contains the actual volumes that ought to be realized.

\subsection*{Derivation of an Algebraic Opinion Pooling Method}

Michael Batty, in his seminal paper "Evolving a Plan: Design and Planning with Complexity", proposes a process of Opinion Pooling for urban design based on his earlier idea of Markov Design Machines. %#TODO reference needed
The typical problem addressed in this book chapter is a recurrent theme in his works pertaining to the human complexity of multi-actor (multi-agent) decision making and finding a satisfactory plan of actions (resource allocation, i.e. where shall we invest our resources?) with respect to a set of sites or investment portfolio objects (called factors in his formulation). 

For convenience, we shall be using Batty's terminology but with a slightly different notation pertaining to our generalized problem. 

There exist a set of $m$ actors/agents and $n$ sites/objects. Each actor has:
\begin{itemize}
    \item a degree of interest $\mathbf{X}=[X_{i,j}]_{m\times n}$, s.t. $\mathbf{X}\mathbf{1}=\mathbf{1}$, i.e. this bipartite matrix is row-stochastic, or that the relative interests of each agent over the $n$ objects add up to exactly 100\%, and
    \item a degree of control $\mathbf{C}=[C_{j',i'}]_{n\times m}$, s.t. $\mathbf{C}\mathbf{1}=\mathbf{1}$, i.e. this bipartite matrix is row-stochastic, or that the relative controls of the $m$ agents over each object add up to exactly 100\%
\end{itemize}
Respectively the two problems will be solved as below:
\begin{itemize}
    \item $\mathbf{P}:=[P_{i,i'}]_{m\times m}=\mathbf{X}\mathbf{C}$: The Markovian interaction probability matrix between agents through their interests and controls over the objects, whose entries represent the relative importance of connections based on the interest \& control connections
    \item $\mathbf{Q}:=[Q_{j',j}]_{n\times n}=\mathbf{C}\mathbf{X}$: The Markovian interaction probability matrix between factors through the interests and controls of the agents, whose entries represent the relative importance of connections based on the control \& interest connections
\end{itemize}
The Plan Design problem can be defined as 'a process of resolving the differences between agents w.r.t. their planned collective investments on the objects given their different interests and controls'. We can define two processes of "conflict resolution" on the primal or the dual problem.

\textbf{The Primal Problem}:
 Consider a negotiation process/game in which the agents collectively redistribute an arbitrary/existing allocation of resources according to the expressed interests and agreed controls. \emph{This is opinion pooling across sites over the distribution of 1 colour amongst actors.} Thus, this form of the problem is only theoretically important but not relevant in this case. It turns out that such a process corresponds to an "ergodic Markov chain" that has a steady-state distribution of resources. The arbitrary initial allocation/distribution of resources is dubbed $\boldsymbol{\alpha}^{(0)}=[\alpha_i]_{1\times m}$, i.e. $\boldsymbol{\alpha}$ is a \emph{row vector}. A Markov chain can be formulated as follows:
 \begin{equation}
 {\alpha}^{(1)}={\alpha}^{(0)}\mathbf{P}\Rightarrow {\alpha}^{(t)}={\alpha}^{(t-1)}\mathbf{P} \Rightarrow {\alpha}^{(t)}={\alpha}^{(0)}\mathbf{P}^t.
 \end{equation}
 
This Markov process is ergodic under relatively simple conditions and thus it can be expected to converge to a steady-state solution characterizing an equilibrium state in which the relative pooling of the categorical (coloured) resources according to all the agents has stabilized as a function of interaction matrix (that reflects relative interests and controls), i.e. in the steady-state: 
\begin{equation}
\lim_{t\rightarrow \infty}\boldsymbol{\alpha}^{(t)}={\alpha}^{(t)}\mathbf{P},
\end{equation}
because, by definition, at the steady state $\boldsymbol{\alpha}^{(t)}=\boldsymbol{\alpha}^{(t-1)}$. Now, if for convenience we denote this steady-state as $\lim_{t \rightarrow \infty} \boldsymbol{\alpha}^{(t)}:=\boldsymbol{\alpha}$, and so, we can rewrite the last equation as below:
\begin{equation}
\boldsymbol{\alpha}=\boldsymbol{\alpha}\mathbf{P}.
\end{equation}
In this form, it is easy to see that $\boldsymbol{\alpha}$ is a left-eigenvector of $\mathbf{P}$ or that $\boldsymbol{\alpha}^T$ a right-eigenvector of $\mathbf{P}^T$ with an eigenvalue equal to $1$. At the same time, this eigenvector must be a stochastic vector (a discrete probability distribution), i.e. $\boldsymbol{\alpha}\mathbf{1}=1$. Now, following an algebraic procedure \cite[~pp.250-252]{nourian_configraphics_2016}, we can find such a stochastic eigenvector using a standard Gaussian [Least Squares] solver. Consider the steady-state solution:
\begin{equation}
\boldsymbol{\alpha}=\boldsymbol{\alpha} \mathbf{P} \Rightarrow \boldsymbol{\alpha} \left (\mathbf{I}-\mathbf{P} \right )=\mathbf{0} \;and\; \boldsymbol{\alpha}\mathbf{1}=1
\end{equation}
We can combine these two equations into one system of linear equations:
\begin{equation}
\boldsymbol{\alpha}\left [ \left (\mathbf{I}_{m\times m}-\mathbf{P}\right )|\mathbf{1}_{m\times 1}\right]=\left [\mathbf{0}_{1\times m}|1 \right] \Rightarrow \left [ \left (\mathbf{I}_{m\times m}-\mathbf{P}\right )|\mathbf{1}_{m\times 1}\right]^T \boldsymbol{\alpha}^T=\left [\mathbf{0}_{1\times m}|1 \right ]^T.
\end{equation}
The latter equation can be solved by in the sense of finding a least squares solution using a standard numerical solver for the $\mathbf{M}\mathbf{x}=\mathbf{a}$ (see the complete notation in the Opinion Pooling Table in the paper) to find this stochastic eigenvector, i.e. 
\begin{equation}
\boldsymbol{\alpha}^T=\arg \min_{\mathbf{x}}{\|\left[ \left (\mathbf{I}_{m\times m}-\mathbf{P}\right )|\mathbf{1}_{m\times 1}\right]^T\mathbf{x}-\left [\mathbf{0}_{1\times m}|1 \right ]^T\|_2}.
\end{equation}

\textbf{The Dual Problem}: consider a negotiation process/game in which an arbitrary/existing allocation of values on factors the factors is collectively pooled by agents according to their designated control and expressed interests. \emph{This is opinion pooling across actors over the distribution of 1 colour amongst sites.} Thus, this is exactly the problem we need to solve. It turns out that such a process corresponds to an ergodic Markov chain that has a steady-state distribution of values. The arbitrary initial allocation/distribution of values is dubbed $\boldsymbol{\beta}^{(0)}=[\beta_j]_{1\times n}$, i.e. $\mathbf{\beta}$ is a \emph{row vector}. The rest of the process is similar to the primal problem and so, for the sake of brevity, only the equations are presented:
A Markov chain can be formulated as follows:
\begin{subequations}
\begin{gather}
{\beta}^{(1)}={\beta}^{(0)}\mathbf{Q} \Rightarrow {\beta}^{(t)}={\beta}^{(t-1)}\mathbf{Q} \Rightarrow {\beta}^{(t)}={\beta}^{(0)}\mathbf{Q}^t\\
\lim_{t \rightarrow \infty} \boldsymbol{\beta}^{(t)}:=\boldsymbol{\beta}\Rightarrow \boldsymbol{\beta}=\boldsymbol{\beta}\mathbf{Q}\\
\boldsymbol{\beta}=\boldsymbol{\beta} \mathbf{Q} \Rightarrow \boldsymbol{\beta} \left (\mathbf{I}-\mathbf{Q} \right )=\mathbf{0} \;and\; \boldsymbol{\beta}\mathbf{1}=1\\
\boldsymbol{\beta}\left [ \left (\mathbf{I}_{n\times n}-\mathbf{Q} \right )|\mathbf{1}_{n\times 1}\right]=\left [\mathbf{0}_{1\times n}|1 \right ] \Rightarrow \left [ \left (\mathbf{I}_{n\times n}-\mathbf{Q}\right )|\mathbf{1}_{n\times 1}\right]^T \boldsymbol{\beta}^T=\left [\mathbf{0}_{1\times n}|1 \right ]^T\\
\boldsymbol{\beta}^T=\arg \min_{\mathbf{y}}{\|\left[\left (\mathbf{I}_{n\times n}-\mathbf{Q}\right )|\mathbf{1}_{n\times 1}\right]^T \mathbf{y}-\left [\mathbf{0}_{1\times n}|1 \right ]^T\|_2}.
\end{gather}
\end{subequations}

%By comparing the two steady-state solutions for the primal and the dual problems we can see that there is a direct correspondence between their information contents: If we take the primal steady-state problem $\boldsymbol{\alpha}=\boldsymbol{\alpha}\mathbf{P}=\boldsymbol{\alpha}\mathbf{X}\mathbf{C}$, multiplying both sides from the right by $\mathbf{X}$ will reveal that: 
%$\boldsymbol{\alpha}\mathbf{X}=\boldsymbol{\alpha}\mathbf{P}\mathbf{X}=\boldsymbol{\alpha}\mathbf{X}\mathbf{C}\mathbf{X}=\boldsymbol{\alpha}\mathbf{X}\mathbf{Q}$, which implies that the vector $\boldsymbol{\alpha}\mathbf{X}$ is a left-eigenvector of $\mathbf{Q}$, however, $\mathbf{Q}$ has only one unique left-eigenvector for the eigenvalue 1, and so we conclude that $\boldsymbol{\alpha}\mathbf{X}=\boldsymbol{\beta}$. which implies that the vector $\boldsymbol{\alpha}\mathbf{X}$ is a left-eigenvector of $\mathbf{Q}$, however, $\mathbf{Q}$ has only one unique left-eigenvector for the eigenvalue 1, and so we conclude that $\boldsymbol{\alpha}\mathbf{X}=\boldsymbol{\beta}$. 
It is straightforward to verify that:
\begin{equation}
\left\{ \begin{matrix}
\boldsymbol{\beta}=\boldsymbol{\alpha}\mathbf{X}\\ 
\boldsymbol{\alpha}=\boldsymbol{\beta}\mathbf{C}
\end{matrix} \right \}.
\end{equation}
    
    From this it is clear that such correspondence not only exists in the two steady states but also during the transient states. In other words, given any distribution of resources over agents or values over the factors, we can find the corresponding dual distribution reciprocally. In particular, given a distribution of resources amongst agents $\boldsymbol{\alpha}^{(t)}$, one can easily find the distributed values of the factors of that iteration time by applying the relative interest of the agents over all factors as captured by the matrix $\mathbf{x}$. This is to say that the RHS of the first equation distributes the resources of agents over the factors and produces the values invested in all factors. Analogously, given a distribution of invested values over the factors one can work out how much each agent has invested their resources by using the matrix $\mathbf{Q}$ (the RHS of the second equation). This also implies that throughout the transient states, the resource and value distributions between two iterations are unequal but get closer to their next iteration values if we feed the next result back into the same pooling process, i.e. make the Markov Chain.
    
\begin{figure}[h]
    \centering
    \includegraphics[width=14 cm]{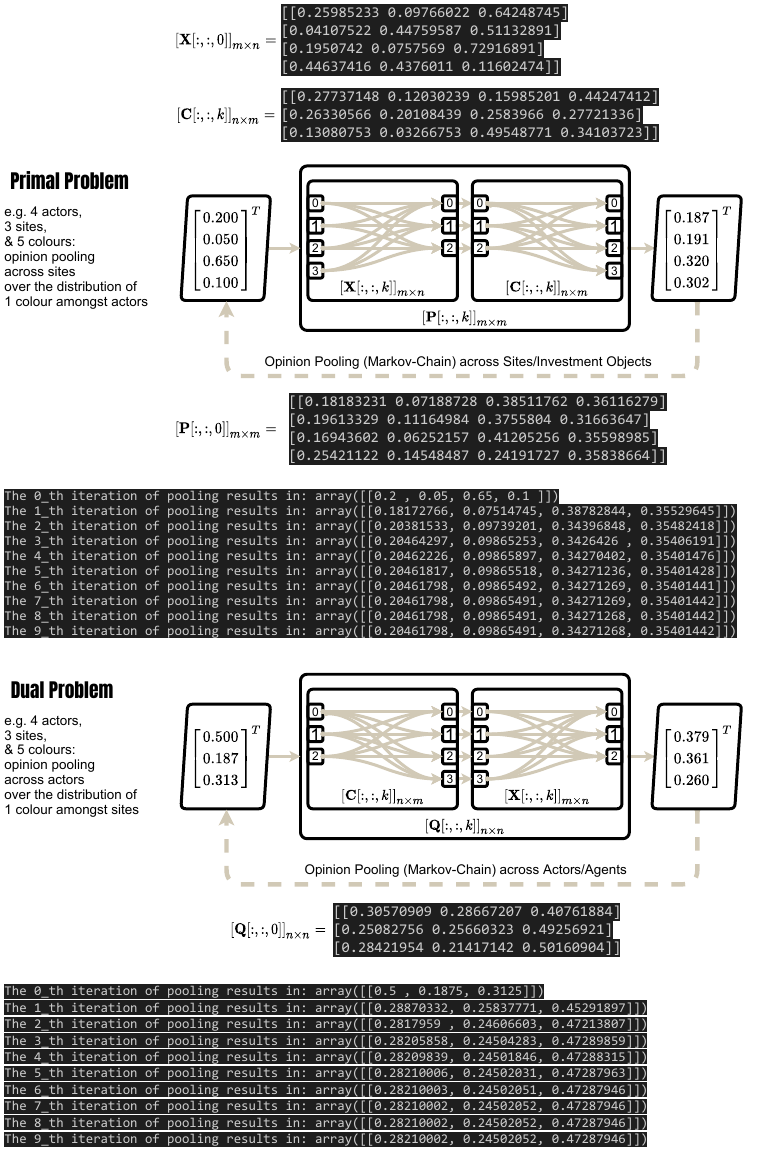}
    \caption{an illustrative example of Iterative Opinion Pooling processes}
    \label{fig:opinion-pooling}
\end{figure}

\subsection*{Derivation of an Algebraic Proportional Fitting Method}

We have condensed this procedure as an algebraic algorithm as follows: the target row-sums and column-sums are iteratively scaled alternatively with column-relativised (ensured to be column stochastic) and row-relativised (ensured to be row stochastic) versions of the original matrix until the series converges, i.e. the norm of the differences between the actual row/column sums with the target sums becomes negligible. 
The row-sums and column-sums are denoted as vectors 
\begin{equation}
\boldsymbol{\rho}_{n\times 1}:=\mathbf{A}_{n\times o}\mathbf{1}_{o\times 1} \;\&\; \boldsymbol{\kappa}_{o\times 1}:=\mathbf{A}_{o \times n}^T \mathbf{1}_{n\times 1},
\end{equation}
reach target values, denoted as vectors 
\begin{equation}
\boldsymbol{\mathfrak{r}}^{(T)}:=[\mathfrak{r}_i^{(T)}]_{n\times 1}\;\&\; \boldsymbol{\mathfrak{c}}^{(T)}:=[\mathfrak{c}_j^{(T)}]_{o\times 1}. 
\end{equation}
The procedure in vectorized (algebraic) form is to iteratively adjust the row-sums and column sums iteratively until the row sums and the column sums get close enough to the target row sums and column sums. Therefore, in a vectorized form, we can write the row-sum adjustment/fitting operation as a combination of relativisation or conversion to a row/column-stochastic matrix and then scaling the rows from the left or scaling the columns from the right, as explained in Algorithm 2 in the paper.

\FloatBarrier

\section*{Game-play}

We have put a web-based prototype of the game to test through three participatory serious gaming workshops. The first two workshops were mainly dedicated to testing the prototype and collecting qualitative feedback on the interface and desired features. The last workshop was to obtain structured test data and collect first-hand feedback from role-players on how real stakeholders could/would interact with the system potentially at massive participation scales. For the sake of brevity, in what follows only the set up and the data collected from the last test-play workshop is discussed.

The case study of the workshop was a former sub-urban industrial district in a historical city in the Netherlands.
Recently the municipality had decided to redevelop the district with the aim to convert it into a high-density residential urban neighbourhood.
This redevelopment process is supposed to add extra \emph{Residential}, \emph{Commercial}, \emph{Cultural}, and \emph{Public} spaces as distinct categories of designated uses (referred to as functions in the architects' jargon) to the district.
These four categorical functionalities, in addition to the \emph{Empty}, comprise the five Colors of the game (regarded as categorical data dubbed $c_k$ in the paper).
There was already a proposal for the redevelopment of the district by a design firm with an idea to preserve the industrial character of the site as much as possible. Accordingly, we divided the district into seven constituent sites closely following its current spatial composition, with the idea to preserve as much as possible the structure of the district: \emph{Skin}, \emph{North Wing}, \emph{Central Area}, \emph{South Wing}, \emph{Southern Yard}, \emph{Central Yard}, and \emph{Kruithuis Yard}.
In the last workshop, five actors played the emblematic roles of \emph{The Mayor}, \emph{The Neighbour}, \emph{The Architect}, \emph{The Inhabitant}, and \emph{The Developer}.

\begin{figure}
    \centering
    \includegraphics[width=13 cm]{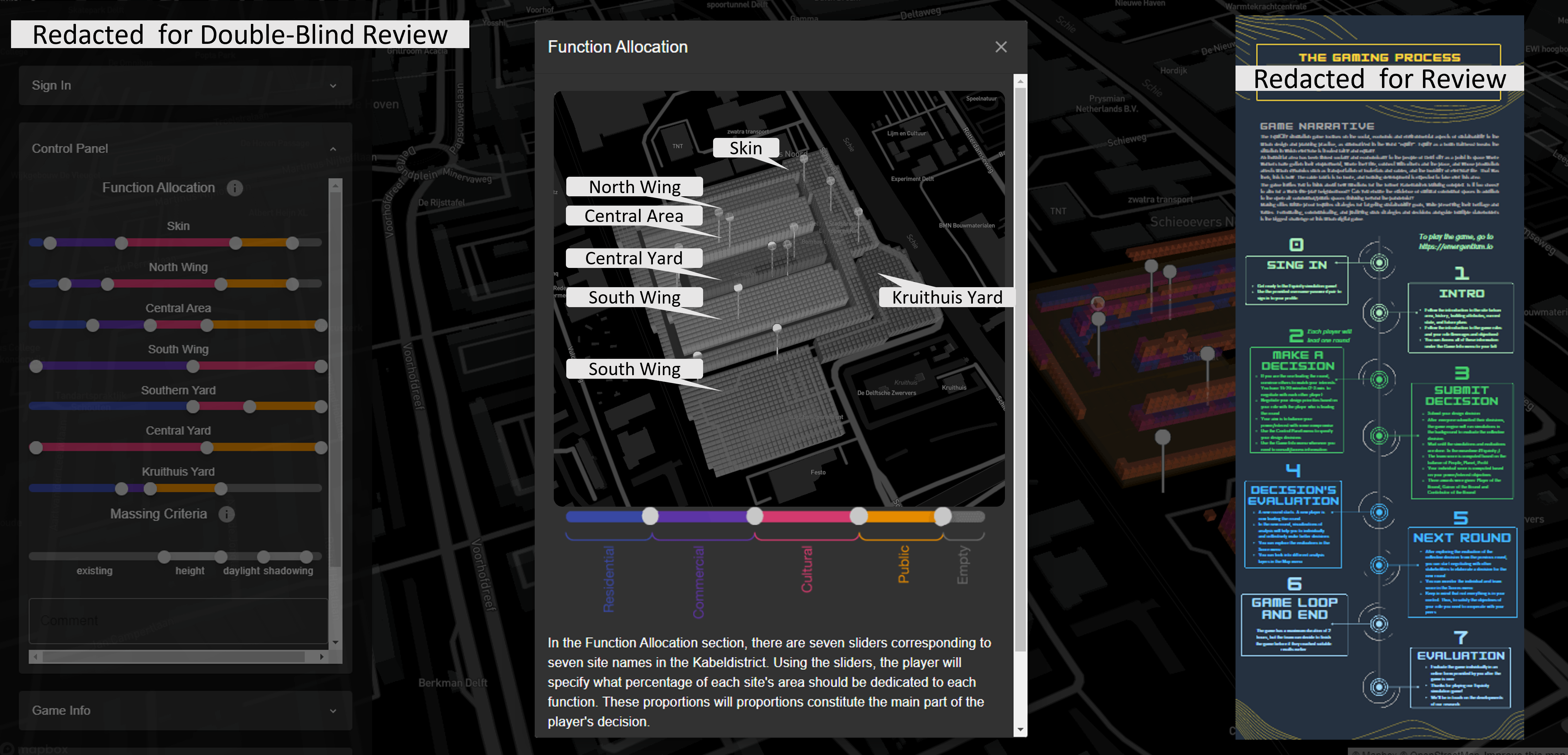}
    \caption{Screenshots of the game interface and its infographic brochure}
    \label{fig:game_interface}
\end{figure}

The gaming workshop was held online; each participant joined a shared video call and accessed the web-based interface using their credentials. They had received the credentials together with an invitation and a brochure explaining the interface, the narrative, and the riles of the game, as shown in Figure \ref{fig:game_interface}. This setting allowed us to record the proceedings of the workshops not only in terms of interaction data but also the verbal conversations (video link removed for double-blind review).

The session took 150 minutes, including a 40-minute introduction to the game, 40 minutes of playing, a 10-minute break, 40 minutes of playing, and finally, 20-minute feedback and reflection.
The participants had no prior knowledge of the game interface.

During this 80 minutes of game-play, the players played three complete rounds.
In each round, one of the players was chosen to be the host (negotiation moderator) of the round randomly.
The host would start the negotiations with other players to persuade them to vote in favor of his/her interests (represented by their interest matrix).
In addition to the interest matrix, players had access to control and power surplus matrices (the latter was called the difference matrix at the time), representing the level of control and mismatch of their interest and control respectively.
Based on the difference (power surplus) matrix, players could understand what areas of interest are outside of their control, so they should rely on other players to support them there to steer the design towards their interest matrix; and what areas of their control they have no interest in so they can use as tokens of power barter in their negotiations.

After the negotiations were concluded (if necessary with a time pressure exerted by the game master), the players would proceed to input their decision (effectively update their interest matrices) about the allocation of space in each site to each color, as well as their assigned weights of importance for the massing quality criteria and submit their decisions in the interactive interface.
When the system ensures that all players have submitted their decisions in a round, the back-end starts to go through the opinion pooling, iterative proportional fitting, massing, and evaluation processes to provide feedback regarding the aggregate outcome of the collective decisions of players in terms of group scores, individual scores, and individual badges. 

The near real-time communication of the outcomes of interest (aggregate quality criteria dubbed as $q_\iota$) marks the end of a round of game-play, after which players can access the scores, badges, and extra info to understand the performance of their decision and contemplate on changing their decisions in the subsequent round accordingly. 
The game interface provides extra information regarding the spatial quality criteria used for massing (in pop out windows). 
This type of information, i.e. the association between the individual decisions and the aggregate outcomes of interest for the group was found to be the most abstract for the players to grapple with in all workshops. We presume that this complex association is what is colloquially known as complexity of design in terms of the difficulty of adjusting the decisions for obtaining a satisfactory balance on measurable outcomes of interest.

\subsection*{Game Master Dashboard}

To assist the game master in assessing the game play process, we put forth a set of statistical workflows that provide insight into how the players are making decisions, how effective these decisions are, and finally, whether the decisions are improving during the game play workshop. To this end, we restate the following assumptions as hypotheses to be tested statistically:

\begin{itemize}
    \item Decision Homogeneity: Decisions are not homogeneous across actors, w.r.t. sites, and colors.
    \item Decision Change: Decisions are changing throughout the rounds.
    \item Discussion Time: Longer rounds will yield a more significant improvement in the decisions.
\end{itemize}

Given that in each round, each actor makes a decision about how much space in each site is allocated to each color, the decisions are formatted as $\mathbf{X}_{\tau \times m\times n \times o}$ where $\tau$ is the counter of discrete time iterators or the number of played \& recorded game-rounds, $m$ is the number of actors, $n$ is the number of sites, $o$ is the number of colors. 
Furthermore, since scores are calculated in each round per each criterion (three criteria in addition to $m$ individual criteria showing the attainment of each individual as the extent to which they have been a gainer in the previous round), the scores are formatted as $\mathbf{O}_{t \times (m+3)}$.
It must be noted that the framework is devised to be scalable for mass-scale participation and the proposed statistical hypothesis testing procedures would eventually require more data to be conclusive. Nevertheless, the procedures are illustrative of the kind of information and insight that the game master could receive to utilize for activating the negotiations.

\subsection*{Decision Homogeneity}
We assume that decisions are not homogeneous across actors, w.r.t. sites, and colors. 
The importance of this assumption is that if the decisions turn out to be homogeneous, then the multi-actor complexity of the game is trivialized.
To test this hypothesis, firstly we check the homogeneity of the data for equal variances using Levene's $W$.
%#TODO Introduce Levene test
\begin{table}[H]
\centering
\begin{tabular}{lr}
\toprule
               Parameter &     Value \\
\midrule
     Test statistics ($W$) &    0.726 \\
 Degrees of freedom ($df$) & 139 \\
                 $p$ value &    .982 \\
\bottomrule
\end{tabular}

\caption{
    \label{tab:homogenity} The result of Levene test for the final workshop game-play data
}
\end{table}

The data-set with three-way design has equal variance with Levene test ($W=0.727, df=139, p=.983$), which fulfills the assumption for ANOVA. 
%Normal ANOVA is used for later comparison.

\begin{table}[H]
    \centering
    \begin{tabular}{lrrrrrr}
\toprule
               Source &   $SS$ & $df$ &    $MS$ &     $F$ &    $p$ & $\eta^2$ \\
\midrule
                Actor &  0.370 &    4 &   0.093 &  23.630 &   .000 &   .252 \\
                 Site &  0.066 &    6 &   0.011 &   2.794 &   .012 &   .056 \\
                Color &  0.665 &    3 &   0.222 &  56.636 &   .000 &   .378 \\
         Actor * Site &  0.286 &   24 &   0.012 &   3.044 &   .000 &   .207 \\
        Actor * Color &  1.274 &   12 &   0.106 &  27.106 &   .000 &   .537 \\
         Site * Color &  0.972 &   18 &   0.054 &  13.793 &   .000 &   .470 \\
 Actor * Site * Color &  1.619 &   72 &   0.022 &   5.742 &   .000 &   .596 \\
             Residual &  1.097 &  280 &   0.004 &     NaN &    NaN &    NaN \\
\bottomrule
\end{tabular}

    \caption{
        \label{tab:ANOVA} The result of three-way ANOVA test for the final workshop game-play data assessing the effect interaction of Actor, Site, and Color
    }
\end{table}

The three-way ANOVA shows that the main effects of Actor, Site, and Color are significant: 
$F_\text{a}(4,280)=23.630$, $p<.001$,$\eta^2=.252$; 
$F_\text{s}(6,280)=2.794$, $p=.012$ ,$\eta^2=.056$; 
$F_\text{c}(3,280)=56.636$, $p<.001$,$\eta^2=.378$. 
The two-way interaction effect between Actor-Site, Actor-Color, Site-Color are all significant: 
$F_\text{a*s}(24,280)=3.044$, $p<.001$,$\eta^2=.207$; 
$F_\text{a*c}(12,280)=27.106$, $p<.001$,$\eta^2=.537$; 
$F_\text{s*c}(18,280)=13.793$, $p<.001$,$\eta^2=.470$. 
And the three-way interaction effect among Actor, Site, and Color is significant: $F_\text{a*s*c}(72,280)=5.742$, $p<.001$,$\eta^2=.596$. 

\begin{figure}[H]
    \centering
    \includegraphics[width=11.25cm]{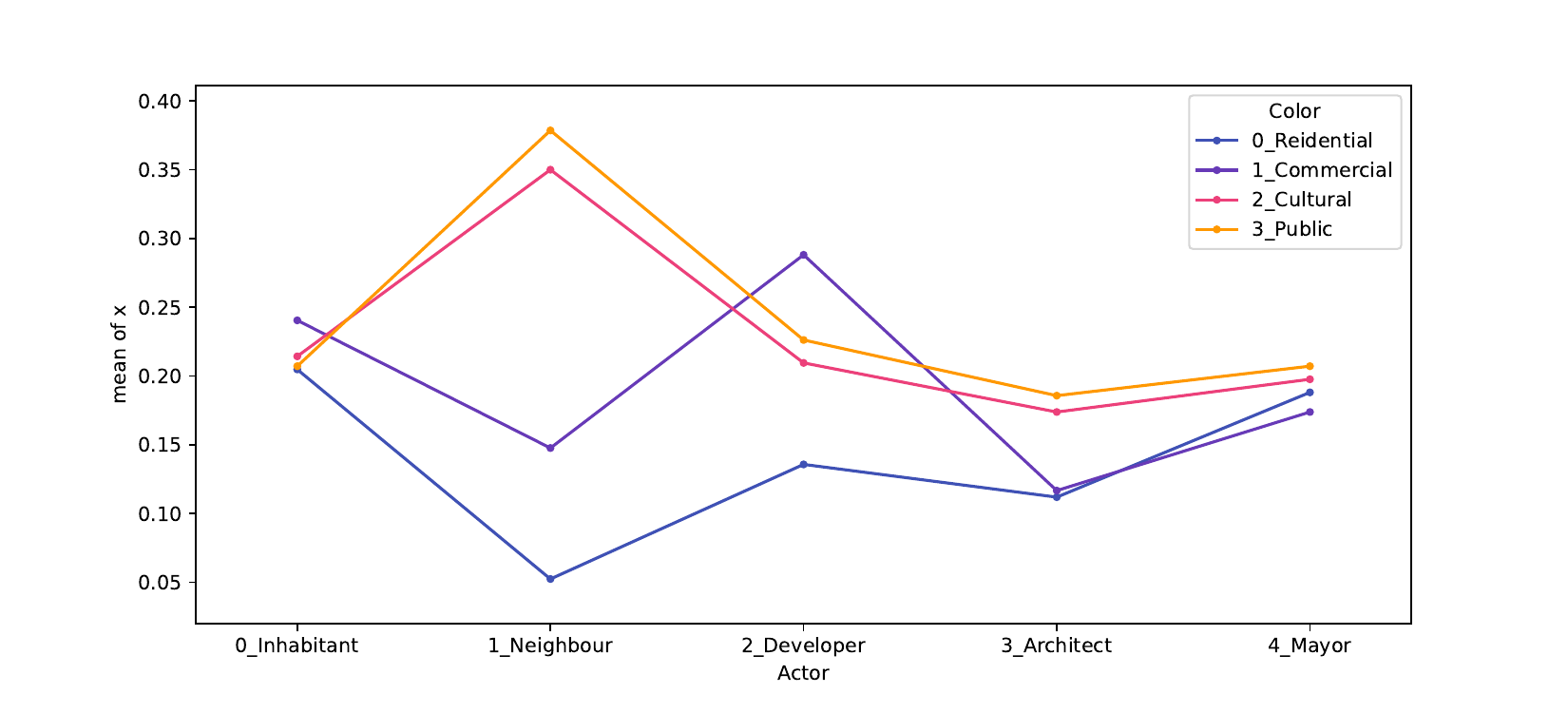}
    \caption{Interaction plot of actors and colors, revealing that amongs actors some have aimed for uniform distributions, (the inhabitant and the mayor) and that the neighhbour seems to have had the sharpest distinction of spatial categories in their treatment of colours}
    \label{fig:interaction_ac}
\end{figure}
\begin{figure}[H]
    \centering
    \includegraphics[width=11.25cm]{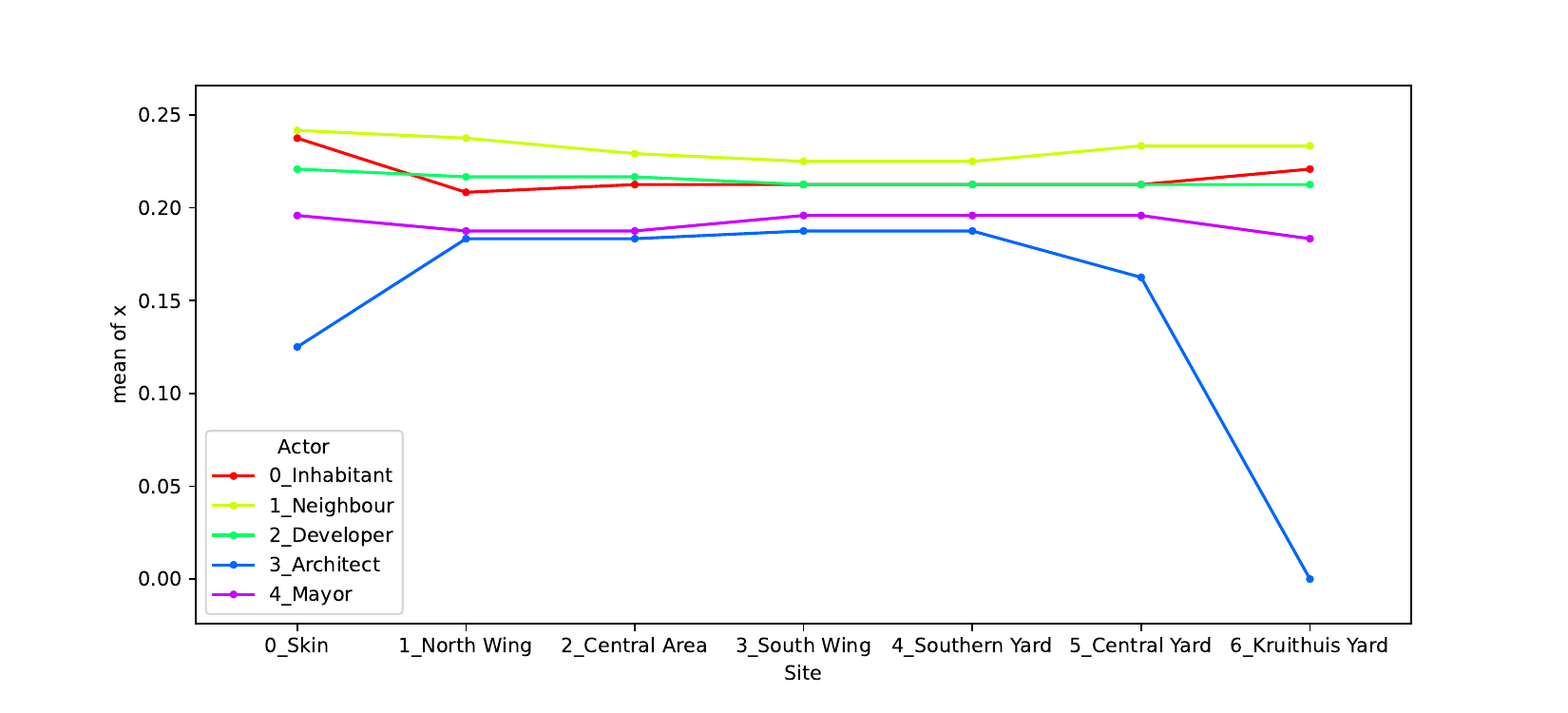}
    \caption{Interaction plot of sites and actors, revealing e.g. that the architect has made the most distinct spatial allocation decisions to the sites, apparently in an attempt to achieve a certain shape for the resultant mass configuration}
    \label{fig:interaction_sa}
\end{figure}
\begin{figure}[H]
    \centering
    \includegraphics[width=11.25cm]{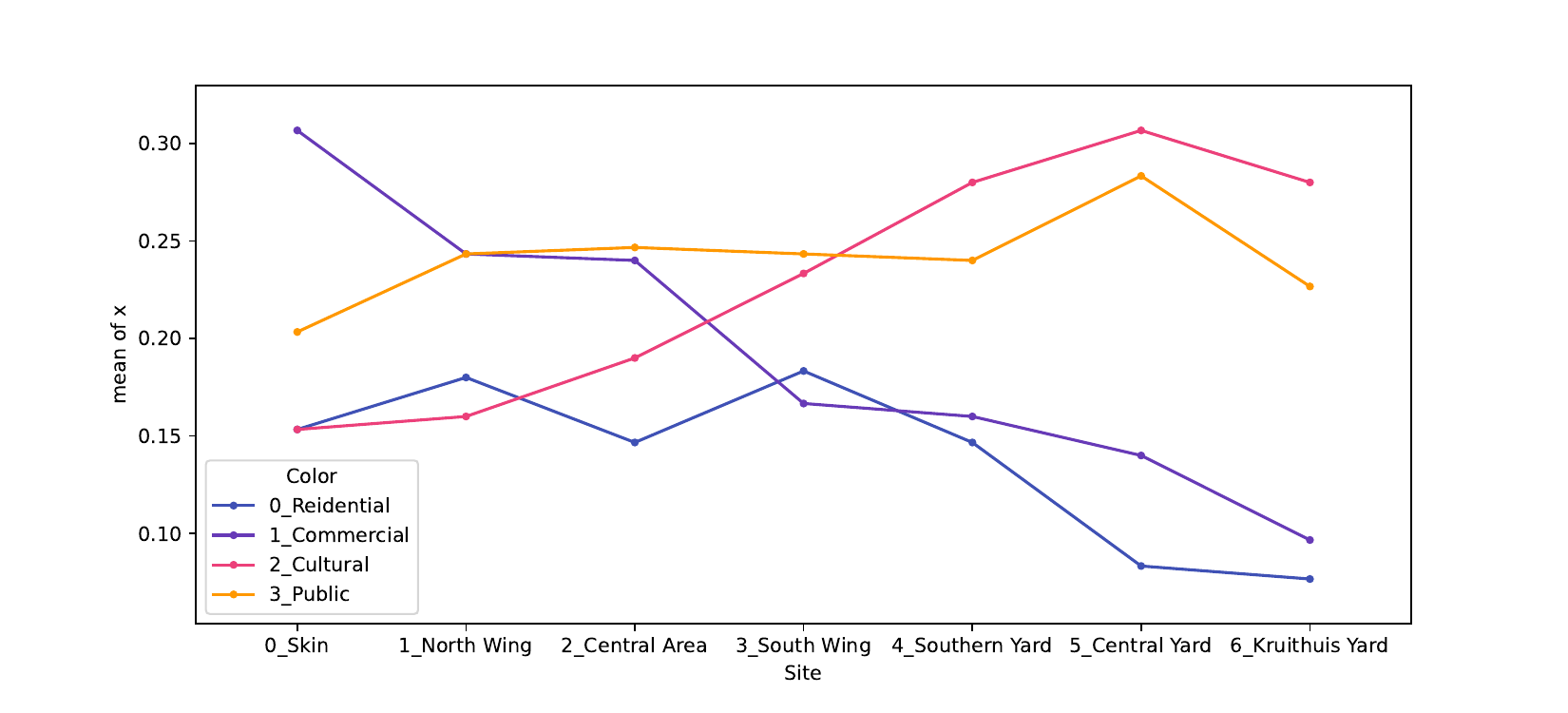}
    \caption{Interaction plot of sites and colors, revelating that the sites have been treated distinctly w.r.t. their colour distributions by all actors throughout the played rounds}
    \label{fig:interaction_sc}
\end{figure}
    
A post-hoc analysis is conducted using Tukey-HSD test to investigate the inter-relationships of Actors, Sites, and Colors.

\begin{table}[H]
    \centering
    \begin{tabular}{llrrrrrrr}
  \toprule
              A &            B &  mean(A) &  mean(B) &   diff &   $SE$ &    $T$ &  $p$-tukey &  hedges \\
  \midrule
   0\_Inhabitant &  3\_Architect &    0.217 &    0.147 &  0.070 &  0.019 &  3.760**   &     .002 &    .578 \\
    1\_Neighbour &  3\_Architect &    0.232 &    0.147 &  0.085 &  0.019 &  4.596***  &     .001 &    .706 \\
    2\_Developer &  3\_Architect &    0.215 &    0.147 &  0.068 &  0.019 &  3.664**   &     .003 &    .563 \\
  \bottomrule
  \end{tabular}
    \caption{
        \label{tab:THSD_A} The result of Tukey-HSD test for Actors. * $p < .05$, ** $p < 0.01$, *** $p < 0.001$ 
    }
\end{table}

The post-hoc Tukey test on Actors showed that actor pairs 0-3, 1-3, and 2-3 were significantly different $p<.005$ while the difference between other actor pairs were insignificant. 
This leads to the conclusion that actors 0,1, and 2 were similar, actor 3 was quite different and actor 4 was lying somewhere in the middle.

% \begin{table}[H]
%     \centering
%     \input{tab/THSD_S.tex}
%     \caption{
%         \label{tab:THSD_S} The result of Tukey-HSD test for Sites. * $p < .05$, ** $p < 0.01$, *** $p < 0.001$ 
%     }
% \end{table}

The post-hoc Tukey test on the Sites showed that the none of the sites were treated significantly different by the decisions. This entails that the treatment of all sites by actors in their decision making is not meaningfully different in this test case. It can be conjectured that the relative unfamiliarity of the actors with the district renders their decisions monotonous w.r.t. the sites.

\begin{table}[H]
    \centering
    \begin{tabular}{llrrrrrrr}
\toprule
A &            B &  mean(A) &  mean(B) &   diff &   $SE$ &    $T$ &  $p$-tukey &  hedges \\
\midrule
 0\_Reidential &  1\_Commercial &    0.139 &    0.193 & -0.055 &  0.016 & -3.394** &    .004 &  -.467 \\
 0\_Reidential &    2\_Cultural &    0.139 &    0.229 & -0.090 &  0.016 & -5.608*** &    .001 &  -.771 \\
 0\_Reidential &      3\_Public &    0.139 &    0.241 & -0.102 &  0.016 & -6.346*** &    .001 &  -.873 \\
 1\_Commercial &      3\_Public &    0.193 &    0.241 & -0.048 &  0.016 & -2.952* &    .018 &  -.406 \\
\bottomrule
\end{tabular}

    \caption{
        \label{tab:THSD_C} The result of Tukey-HSD test for Colors. * $p < .05$, ** $p < 0.01$, *** $p < 0.001$ 
    }
\end{table}
The post-hoc Tukey test on Colors showed that color pairs 0-1, 0-2, and 0-3 were significantly different with $p<.005$, color pair 1-3 had also a relatively significant difference with $p<.05$, while the difference between other color pairs were insignificant. This entails that the most pairs of colored resources have been treated differently by the actors.

\subsection*{Decision Change}

In this hypothesis-testing procedure, we want to check whether the decisions of the players are changing [significantly] as the game progresses.
Therefore we formulate a hypothesis as "Decisions are changing throughout the rounds." 
Note that for this change to convey a sense of progress, we need to have a guage of convergence or a single metric, as to which we could consider the weighted product of all aggregate quality criteria as a single score. 
While this sense of progress is shown in the game dashboard already to all players, the one discussed here is merely about the ANOVA in between the axes of rounds of the game and the average colour percentages over these rounds, as the main objects of interest in the game illustrating the categorical investments. 
Note that there is already a given district level distribution of colours that we refer to as a programme of requirements which is being used as the basis for the iterative proportional fitting of all proposed colours to site allocations. 
These colour distributions could eventually also be compared to that PoR. 
However, since this PoR was not explained to the players extensively we cannot expect to make a meaningful conclusion from such a comparison in this case.

\begin{table}[H]
    \centering
    \begin{tabular}{rrr}
\toprule
 Round &   mean &  standard deviation \\
\midrule
     0 &  0.201 &               0.109 \\
     1 &  0.196 &               0.131 \\
     2 &  0.203 &               0.127 \\
\bottomrule
\end{tabular}

    \caption{
        \label{tab:round_mean_std} Mean and Standard Deviation over the decisions of each round.
    }
\end{table}

\begin{table}[H]
    \centering
    \begin{tabular}{lr}
\toprule
               Parameter &    Value \\
\midrule
     Test statistics ($W$) &   0.467 \\
 Degrees of freedom ($df$) &  11 \\
                 $p$ value &   .141 \\
\bottomrule
\end{tabular}

    \caption{
        \label{tab:round_levene} The result of levene to check the assumption of ANOVA
    }
\end{table}

The data-set with two-way design of round-color has equal variance with Levene test ($W=1.467$, $df=11$, $p=.1412$). Normal ANOVA is used for later comparison.

\begin{table}[H]
    \centering
    \begin{tabular}{lrrrrrr}
\toprule
        Source &   $SS$ & $df$ &    $MS$ &     $F$ &    $p$ & $\eta^2$ \\
\midrule
         Round &  0.004 &    2 &  0.002 &   0.131 &   .877 &   .001 \\
         Color &  0.665 &    3 &  0.222 &  16.241 &   .000 &   .107 \\
 Round * Color &  0.108 &    6 &  0.018 &   1.321 &   .246 &   .019 \\
      Residual &  5.573 &  408 &  0.014 &     NaN &    NaN &    NaN \\
\bottomrule
\end{tabular}

    \caption{
        \label{tab:ANOVA_RC} The result of two-way ANOVA test for the final workshop decisions across rounds; assessing the effect interaction of Round and Color
    }
\end{table}

The two-way ANOVA shows that the main effects of Round is not significant: $F_\text{r}(2,408)=0.131$, $p>.05$,$\eta^2=.001$; the main effect of Color is significant: $F_\text{c}(3,408)=16.241$, $p<.001$,$\eta^2=.107$. The interaction effect between color and round is not significant: $F_\text{r*c}(6,408)=1.321$, $p>.05$,$\eta^2=.019$.

\begin{figure}[H]
    \centering
    \includegraphics[width=12cm]{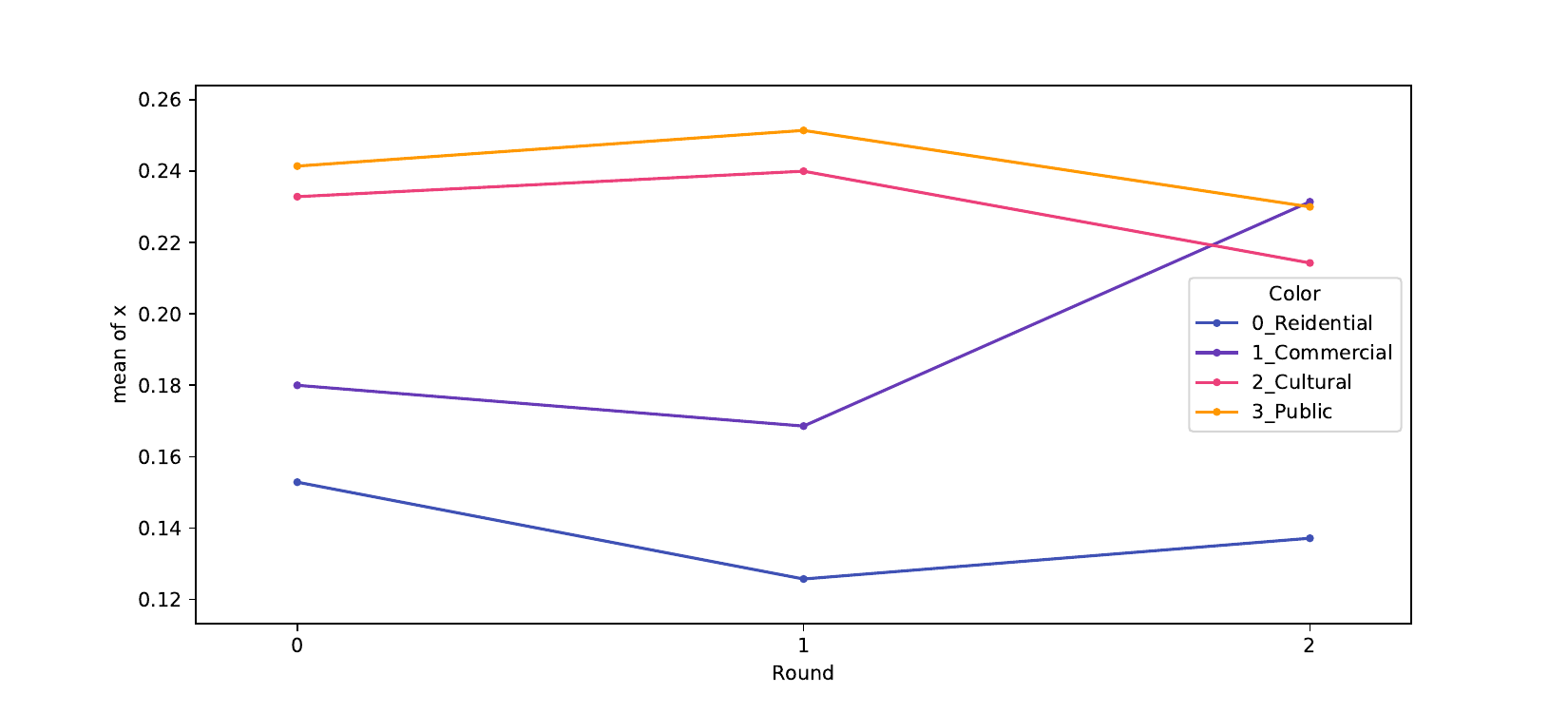}
    \caption{Interaction plot of rounds and colours, revealing that the average percentages of colours allocated to the whole district change radically at least in the last round, with a considerably higher percentage of commercial space allocations}
    \label{fig:interaction_rc}
\end{figure}

% \begin{figure}[H]
%     \centering
%     \includegraphics[width=12cm]{img/interaction_SR.pdf}
%     \caption{Interaction plot of sites and rounds}
%     \label{fig:interaction_sr}
% \end{figure}

\begin{figure}[H]
    \centering
    \includegraphics[width=9.5cm]{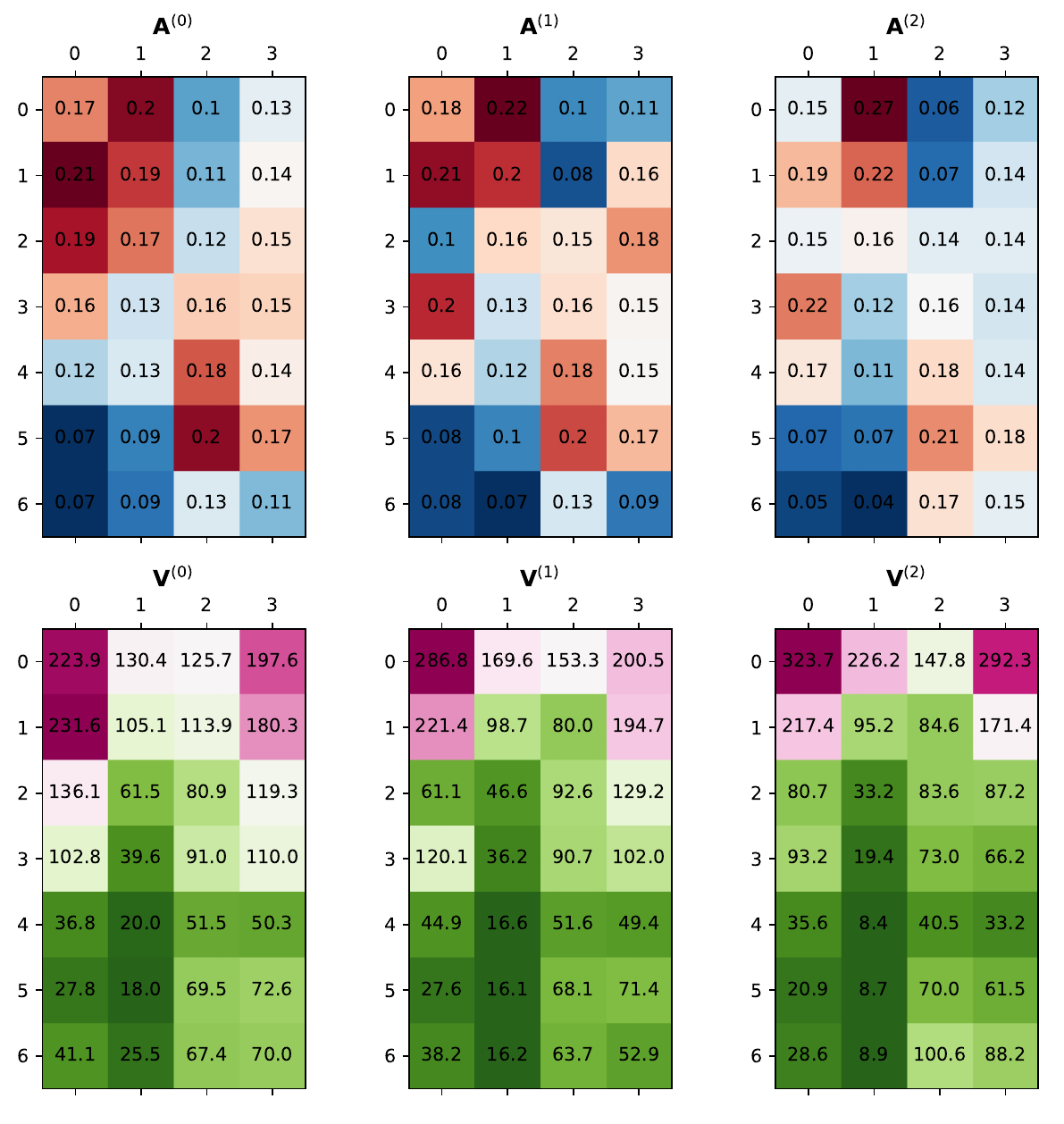}
    \caption{Tensor plots of $\mathbf{A}^{(t)}_{n \times o}$ and $\mathbf{V}^{(t)}_{n \times o}$ during the three rounds of the final workshop, revealing that there exists some level of coherence amongst the decision tensors throughout the rounds and yet the tensors are obviously pointing into different directions, hence the exploratory relevance of the game}
    \label{fig:tensor_plots}
\end{figure}

The interaction patterns of Round-Color (see Figure \ref{fig:interaction_rc}) shows that different colors are distributed equally for each round, showing that no significant changes have occurred along the game-play rounds. These insignificant variations in how spaces in each site are allocated to each color, are more visible in Figure \ref{fig:tensor_plots}. 

\FloatBarrier

\subsection*{Discussion Time}

\begin{figure}[H]
    \centering
    \includegraphics[width=12cm]{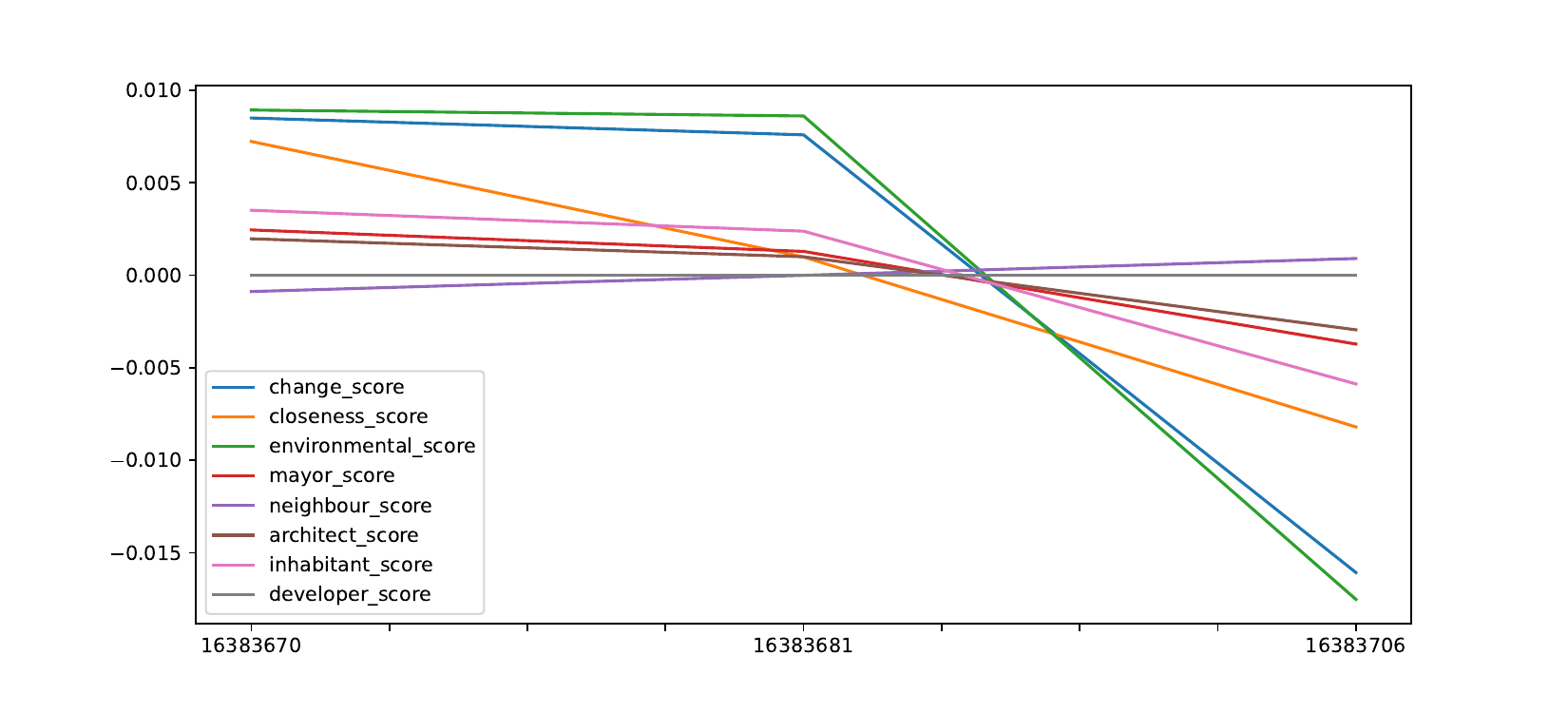}
    \caption{group and individual scores at the end of rounds, revealing that the neighbour and the developer are the only actors who have consistently improved their gains as to their initial agenda cards; while the aggregate sustainability scores have dropped in the last round, calling for further attention to the complexity of the goal-oriented spatial design}
    \label{fig:score_time}
\end{figure}

Finally we looked into whether longer discussions in the rounds could yield a significantly higher score. 
To check this we propose to compute the Pearson correlation between the time spent in each round and the difference in the score of the round compared to previous round. 
In this test we consider all of the group scores ($\mathbf{q}$) and individual scores (gains). 
Nonetheless, since in the final workshop the players have only played three rounds, there would be only two samples for the correlation. 
Therefore, for this case we refer to Figure \ref{fig:score_time} to conclude that the difference between the scores of the first and second round is negligible in most of the scores except closeness score. 
Subsequently, the scores in round three have dropped compared to the second round. 

\FloatBarrier

\section*{Data Availability}
The datasets generated and/or analysed during the current study are available in the EquiCityData repository \\ \href{https://github.com/shervinazadi/EquiCity_Data}{https://github.com/shervinazadi/EquiCity\_Data}.

\section*{Author contributions statement}

%PZN \& SAZ conceived the work, devised the mathematical process, and wrote the first prototypes of the game engine. PZN, SAZ \& APR acquired the funding.  PZN \$ SRZ wrote the first manuscript. SAZ \& NAZ developed the Front-End interface of the game. SAZ \& PZN developed the Back-End of the game; SAZ integrated, developed, and generalized the platform. NBI and SAZ conducted the statistical validation tests for devising the Game Master's dashboard and analysed the results. BDA, SAZ, \& PZN gamified the mathematical design process together. 
Contributions stated according to the \href{https://casrai.org/credit/g}{CRediT – Contributor Roles Taxonomy}:

\textbf{PZN}: Conceptualization, Methodology, Formal analysis, Software, Visualization, Writing- original draft preparation, Funding acquisition, Investigation, Supervision, Project administration, Validation. 
\textbf{SAZ}: Conceptualization, Methodology, Formal Analysis, Software, Data curation, Visualization, Funding Acquisition, Resources, Writing – review \& editing. 
\textbf{NBI}: Validation, Formal analysis, Investigation, Writing – review \& editing. 
\textbf{BDA}: Conceptualization, Data curation, Investigation.
\textbf{NAZ}: Software, Resources. 
\textbf{SRZ}: Conceptualization, Writing- original draft preparation
\textbf{APR}: Conceptualization, Supervision, Funding acquisition.
All authors reviewed the manuscript.
\bibliography{EquiCity_Nature}